\documentclass[aps,pra,twoside,twocolumn,superscriptaddress,floatfix,10pt]{revtex4-2}

\usepackage{preamble_file}
\usepackage{nomencl}
\makenomenclature
\usepackage{etoolbox}
\renewcommand\nomgroup[1]{%
  \item[\bfseries
  \ifstrequal{#1}{B}{Roman Symbols}{%
  \ifstrequal{#1}{C}{Greek Symbols}{%
  \ifstrequal{#1}{A}{Indices}{}}}%
]}
\usepackage{siunitx}
\usepackage{collcell}
\newcolumntype{s}{>{\collectcell\unit}c<{\endcollectcell}}
\newcommand{\nomunit}[1]{%
\renewcommand{\nomentryend}{\hspace*{\fill}\si{#1}}}

\begin{document}

\author{Franziska Kilchert} \affiliation{German Aerospace Center,
  Wilhelm-Runge-Stra{\ss}e 10, 89081 Ulm, Germany}
\affiliation{Helmholtz Institute Ulm, Helmholtzstra{\ss}e 11, 89081
  Ulm, Germany}

\author{Max Schammer} \affiliation{German Aerospace Center,
  Wilhelm-Runge-Stra{\ss}e 10, 89081 Ulm, Germany}
\affiliation{Helmholtz Institute Ulm, Helmholtzstra{\ss}e 11, 89081
  Ulm, Germany}

\author{Arnulf Latz} \affiliation{German Aerospace Center,
  Wilhelm-Runge-Stra{\ss}e 10, 89081 Ulm, Germany}
\affiliation{Helmholtz Institute Ulm, Helmholtzstra{\ss}e 11, 89081
  Ulm, Germany} \affiliation{Universit\"at Ulm, Albert-Einstein-Allee
  47, 89081 Ulm, Germany}

\author{Birger Horstmann} \email{birger.horstmann@dlr.de}
\affiliation{German Aerospace Center, Wilhelm-Runge-Stra{\ss}e 10, 89081 Ulm, Germany} \affiliation{Helmholtz Institute Ulm,
  Helmholtzstra{\ss}e 11, 89081 Ulm, Germany}
\affiliation{Universit\"at Ulm, Albert-Einstein-Allee 47, 89081 Ulm,
  Germany}
\allowdisplaybreaks

\title{Silicon Nanowires as Anodes for Lithium-Ion Batteries: Full Cell Modeling}

\allowdisplaybreaks

\begin{abstract}
Silicon (Si) anodes attract a lot of research attention for their potential to enable high energy density lithium-ion batteries (LIBs). Many studies focus on nanostructured Si anodes to counteract deterioration. In this work, we model LIBs with Si nanowire (NW) anodes in combination with an ionic liquid (IL) electrolyte. On the anode side, we allow for elastic deformations to reflect the large volumetric changes of Si. With physics-based continuum modeling we can provide insight into usually hardly accessible quantities like the stress distribution in the active material. For the IL electrolyte, our thermodynamically consistent transport theory includes convection as relevant transport mechanism. We present our volume-averaged 1d+1d framework and perform parameter studies to investigate the influence of the Si anode morphology on the cell performance. Our findings highlight the importance of incorporating the volumetric expansion of Si in physics-based simulations. Even for nanostructured anodes -- which are said to be beneficial concerning the stresses -- the expansion influences the achievable capacity of the cell. Accounting for enough pore space is important for efficient active material usage.
\end{abstract}

\maketitle
\tableofcontents

\section{Introduction}
\label{sec:Introduction}

Lithium-ion batteries (LIBs) are the state-of-the-art rechargeable energy storage technology for mobile and portable applications. Nevertheless, the ever increasing demand for high energy density solutions constitutes an enduring research stimulus for improving the cell components. Novel materials for next-generation LIBs include Silicon (Si) as active anode material and ionic liquids (ILs) as electrolyte.

Because of its superior theoretical energy density (\qty{3579}{\milli \ampere \hour \per \g} or \qty{2190}{\milli \ampere \hour \per \cm \cubed})\cite{Obrovac2004,Obrovac2007a} compared to the commonly used graphite (\qty{372}{\milli \ampere \hour \per \g} or \qty{837}{\milli \ampere \hour \per \cm \cubed}),\cite{Winter1998} Si is a promising anode material. However, this benefit comes at the cost of large volumetric expansion of Si during lithiation of up to 280\% (compared to less than 10\% for graphite).\cite{Obrovac2004,Winter1998} Together with the relatively slow diffusion of lithium (Li) in Si this leads to high mechanical stresses and consequently fast deterioration of the anode.\cite{Soni2012,Kolzenberg2022a} This challenge is often circumvented by using composite anodes where only a small fraction of the active material is Si.\cite{Zilberman2019,Li2019,Ansean2020,Chen2022} To make use of the high capacity that pure Si anodes provide, another approach is to apply Si nanostructures.\cite{Szczech2011,Wu2012,Rahman2016,Kolzenberg2022a} Studies have shown that below a certain diameter threshold Si is less prone to fracturing.\cite{Cheng2010,Deshpande2010,Ryu2011,Liu2012a} Here, especially Si NWs have sparked great research interest.\cite{Kennedy2016,Zhou2019,Chan2008,Cui2009,Zamfir2013,Yang2020} 

Theoretical modeling and simulation of full cells and cell components can give insights into processes that are otherwise hardly accessible and thus support the development of such next-generation LIBs. Various types of models have been applied to study different aspects of Si anodes. Phase-field models are used to investigate the phase separation and mechanical stresses using finite element method.\cite{Chen2014,Xie2015,Zuo2015,Gao2016,Castelli2021} Many single particle models include the important mechanical effects of Si lithiation.\cite{Deshpande2010,Sikha2014,De2014,Yang2014,Kolzenberg2022a,Koebbing2023} For example, Verma \emph{et al.} study the Si lithiation in nanospheres and nanorods using a single particle model.\cite{Verma2019} Given the wide range of literature values for the solid diffusion coefficient of Li in Si and the rate constant they propose individual values for each Si particle geometry based on comparison with experiments. Physics-based volume-averaged simulations focus often on Si composite electrodes\cite{Chandrasekaran2011,Sturm2019,Lory2020,Chen2022} or apply the models to half-cells with standard electrolytes\cite{Wang2017,Appiah2019} only. However, Chandrasekaran and Fuller already highlight the importance of electrode porosity with Si as active material in the anode.\cite{Chandrasekaran2011}

\begin{figure*}[tb]
	\centering
	\includegraphics[width=0.95\textwidth]{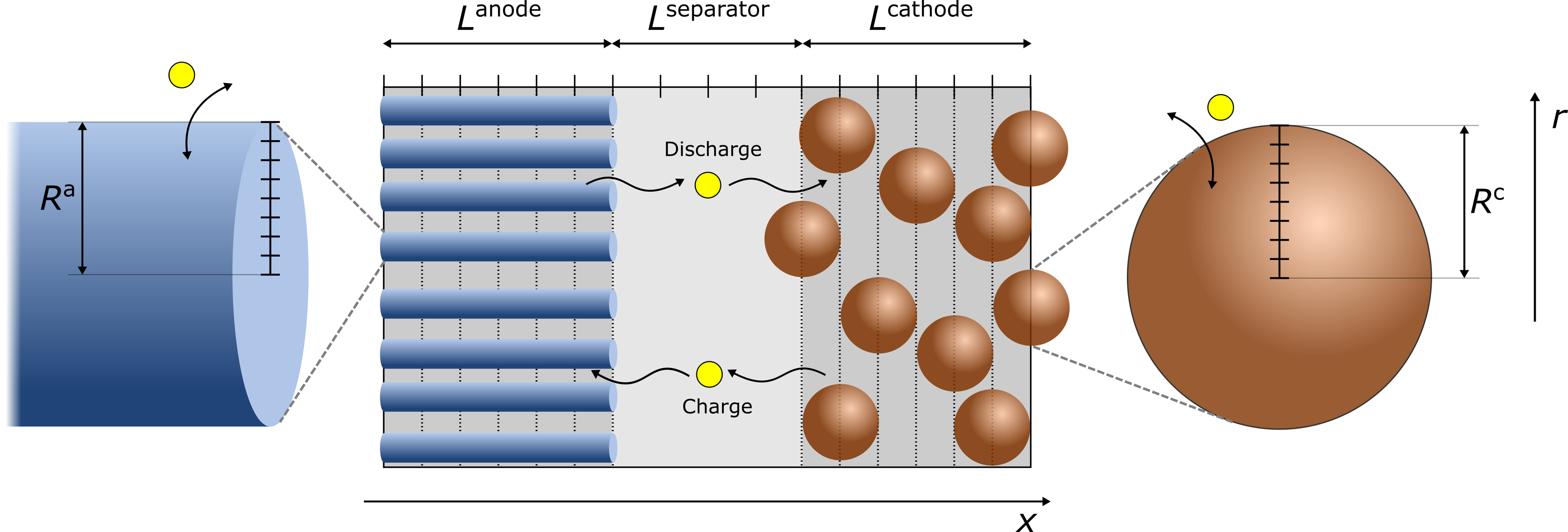}
	\caption{Schematic of the 1d+1d full cell model. Li-ion transport is modeled in $\xcoord$-dimension throughout the cell and in $\rcoord$-dimension within the active electrode particles.}
	\label{fig:CellScheme}
\end{figure*}

ILs are compatible with low voltage electrodes such as Si anodes and can help reduce the SEI-related degradation of Si.\cite{MacFarlane2014,Baranchugov2007,Tang2022} Thus, significant research goes into combining these favorable material classes.\cite{Chakrapani2011,Kim2017,Kerr2017,Stokes2020,Domi2022,Karimi2023} To describe transport in IL electrolytes, important features of highly concentrated electrolytes have to be captured correctly. Early continuum theories already combine the different scales of bulk and interfaces but cannot resolve detailed transport processes.\cite{Bazant2011,Yochelis2015,Gavish2016,Gavish2017,Bier2017} Monroe \emph{et al.} include important aspects of highly concentrated electrolytes in their modeling equations like solute volume effects and stresses due to local pressure.\cite{Monroe2013,LiuJ2014,Goyal2017}

In this work, we use physics-based continuum modeling to bring together nanostructured Si anodes and IL electrolytes in one consistent full cell model. We combine our novel transport theory for highly concentrated electrolytes\cite{Schammer2020theory} with the recently developed chemo-mechanically coupled anode model\cite{Castelli2021,Kolzenberg2022a}. Both theories are derived consistently from non-equilibrium thermodynamics by modeling the free energy and incorporate the crucial aspects of the respective cell component. The electrolyte theory includes convection as non-negligible transport mechanism and correctly reflects the unique interfacial behavior of ILs.\cite{Hoffmann2018,Schammer2021role} On the anode side, the chemo-mechanical coupling results from concentration-dependent reversible deformations.\cite{Kolzenberg2022a} Allowing for these elastic deformations is still important for nanostructured Si anodes due to the large volumetric changes.

We structure this manuscript into two main parts. First, in \cref{sec:Model}, we describe the modeling framework in detail. Second, in \cref{sec:Results}, we apply our model to  secondary battery cells containing amorphous Si nanowires (NW) as anode, IL electrolytes, and a standard NMC cathode. Furthermore, we study the effects of the anode geometry and transport parameters on the cell performance.

\section{Model}
\label{sec:Model}

In this theory chapter, we outline our modeling framework. Our continuum transport theory is based on thermodynamics and focuses on modeling the free energy density. To account for the mechanical aspects of the anode and for the IL-based liquid electrolyte in our holistic description, we combine two separate frameworks, which were previously published by some of the authors. First, our description of the electrolyte is based on the transport theory presented in Refs.~\citenum{Latz2015,Schammer2020theory,Schammer2021role}. Second, we account for the large volumetric changes in the Si-based anode material by applying the chemo-mechanical model for active particles presented in Refs.~\citenum{Castelli2021,Kolzenberg2022a}. Altogether, we couple the two descriptions of the solid and liquid phases via an electrochemical reaction rate.

We structure this modeling chapter as follows. First, in \cref{sec:set_up_1D}, we present our set-up of the battery. Next, in \cref{sec:Theory}, we briefly summarize the major aspects of both theories and state the modeling equations. A more detailed discussion is presented in \cref*{sec:SI_theory} of the ESI, and can be found in the literature.\cite{Schammer2020theory,Schammer2021role,Castelli2021,Kolzenberg2022a} Finally, in \cref{sec:Parameterization}, we discuss the parameterization of the baseline cell.

\begin{table*}[!htb]
\centering
\begin{tabular}{ccc} 
    \toprule
    Anode & Electrolyte & Cathode \\
    \midrule
    $\displaystyle \frac{\partial}{\partial \Time} \theta^\mathrm{a}_\mathrm{Li} = \frac{1}{\Radius_0^\mathrm{a}} \frac{\partial}{\partial \Radius_0^\mathrm{a}} \left[ \Radius_0^\mathrm{a} \frac{\diffcoeff^\mathrm{a}_\mathrm{Li}}{\Rconst \Temp} \soc^\mathrm{a}_\mathrm{Li} (1 - \soc^\mathrm{a}_\mathrm{Li}) \frac{\partial \chempot_\mathrm{Li}}{\partial \Radius_0^\mathrm{a}} \right]$
    & 
    $\displaystyle \frac{\partial}{\partial \Time} \porosity \conc^\mathrm{e}_\mathrm{Li} = - \frac{\partial}{\partial \xcoord} \Nflux_\mathrm{Li} - \frac{\partial}{\partial \xcoord} \porosity \conc^\mathrm{e}_\mathrm{Li} \vel + \source_\mathrm{Li}$ 
    & 
    $\displaystyle \frac{\partial}{\partial \Time} \soc^\mathrm{c}_\mathrm{Li} = \frac{1}{{\Radius^\mathrm{c}}^2} \frac{\partial}{\partial \Radius^\mathrm{c}} {\Radius^\mathrm{c}}^2 \diffcoeff^\mathrm{c}_\mathrm{Li} \frac{\partial \soc^\mathrm{c}_\mathrm{Li}}{\partial \Radius^\mathrm{c}}$
    \\
    $\displaystyle 0 = \frac{\partial \FirstPiola_\mathrm{rev,r}}{\partial \Radius_0^\mathrm{a}} + \frac{1}{\Radius_0^\mathrm{a}} \left( \FirstPiola_\mathrm{rev,r} - \FirstPiola_\mathrm{rev,\upphi} \right)$
    & 
    $\displaystyle 0 = - \frac{\partial}{\partial \xcoord} \Jflux + \Faraday \source_\mathrm{Li}$ 
    & 
    \\
    & 
    $\displaystyle 0 = - \frac{\partial}{\partial \xcoord} \porosity \vel - \Tilde{\Tilde{\pmv}}_\mathrm{Li} \frac{\partial}{\partial \xcoord} \Nflux_\mathrm{Li} - \frac{\Tilde{\pmv}_\mathrm{cat}}{\Faraday} \frac{\mass_\mathrm{an}}{\mass_\mathrm{IL}} \frac{\partial}{\partial \xcoord} \Jflux + \pmv_\mathrm{Li} \source_\mathrm{Li}$ 
    & 
    \\
    \bottomrule
\end{tabular}
\caption{System of differential equations that describes the transport in the anode, the cathode and the electrolyte. Three species are considered in the electrolyte (anion, cation, Li). See \cref{eq:LiFluxElyte,eq:ElCurrDensElyte} for the definition of the Li flux density and the electric current density in the electrolyte. Cylindrical coordinates are used on the anode side, spherical coordinates on the cathode side, and Cartesian coordinates for the electrolyte.}
\label{tab:SystemOfEquations}
\end{table*}

\subsection{1d+1d Set-up of the Battery}
\label{sec:set_up_1D}

In this section, we present our one-dimensional (1d) set-up of the LIB with an IL electrolyte and a Si-based anode. \Cref{fig:CellScheme} shows a schematic of the full cell model with nanowires (NWs) on the anode side in cylindrical geometry and spherical cathode particles. We model the transport of Li-ions throughout the cell in $\xcoord$-direction and separate the battery cell into three distinct regions. The left region comprises the anode, next to the separator in the middle, and the cathode to the right. In addition, we model the active material of the electrodes in radial $\rcoord$-direction, see the cylindrical Si NWs of the anode on the left side and the spherical particles of the cathode on the right side. Altogether, this yields a 1d+1d model of the battery cell. 

The three cell components (anode, separator and cathode) have a porous structure, into which the liquid electrolyte is immersed. Thus, they constitute superpositions of two continuous phases, the active, solid material (s) and the liquid electrolyte (e). We address this complex morphology using porous electrode theory\cite{Newman1975} and discretize the two dimensions ($\xcoord$ and $\rcoord$) into voxels ($\xcoord_1$, ..., $\xcoord_\mathrm{n}$ and $\rcoord_1$, ..., $\rcoord_\mathrm{m}$). The main descriptor of the morphology is the porosity $\porosity$ of the electrodes and separator, which is defined as the relative amount of pore space which is filled with electrolyte. Since in our model the anode is made of Si NWs, which undergo large volume changes, the porosity of the anode is constantly adjusted according to the current radius of the Si NWs. In addition, we phenomenologically account for the tortuosity of the cell components via a Bruggeman coefficient of $\Brugge = 1.5$. Based on this, we describe the transport via volume-averaged quantities. The complex interplay of radius and pore space is investigated in \cref{sec:Param_Study}.

\subsection{Theory}
\label{sec:Theory}

In this section, we summarize the two continuum models for the electrolyte transport and for the mechanical deformation of the Si NWs, on which our theoretical description of the complete cell is based.  

Both continuum models are derived from the same underlying framework of rational thermodynamics (RT).\cite{Schammer2020theory,Schammer2021role,Castelli2021,Kolzenberg2022a} RT is based on rigorous physical assumptions, \emph{e.g.} universal balancing laws, and provides a description of non-equilibrium thermodynamics based on constitutive equations. These take the form of thermodynamic derivatives of the Helmholtz free energy $\freeenergy = \int \dens\freeenergydens \mathrm{d}V$, and are consistent with the thermodynamic laws. As consequence, the focal quantity in this framework is the Helmholtz free energy density of the system, $\dens\freeenergydens$, which casts the general framework to specific materials. \Cref{tab:SystemOfEquations} contains the final (isothermal) set of differential algebraic equations, which describes transport in the battery.

In the following, we briefly discuss the theoretical description of the three main components (electrolyte, anode and cathode), our coupling between the liquid phase and the solid phase via a Butler-Volmer ansatz, and the boundary conditions. A more in-depth derivation of the theories can be found in \cref*{sec:SI_theory} in the ESI.

First, we discuss the electrolyte. Here, we focus on IL-based electrolytes of the form "Li-salt + IL with common anion". Hence, we restrict our general description (applicable to any number of electrolyte species) to three ionic components, which are present in the liquid. These are the Li-ions, the cation of the IL, and the common anion. We account for all relevant transport mechanisms, \emph{i.e.} convection (which plays an important role in concentrated electrolytes), diffusion, and migration. Altogether, the theory results in three electrolyte transport equations, \emph{cf.} \cref{tab:SystemOfEquations}. Here, $\conc_\mathrm{Li}^\mathrm{e}$ is the Li-ion concentration in the electrolyte, $\vel$ the center-of-mass based convection velocity, $\mass_\alpha$ the molar mass of species $\alpha$ and $\source_\mathrm{Li}$ are source terms for Li. We refer to the list of symbols for a detailed explanation of the nomenclature. Furthermore, the flux density of the Li-ions $\Nflux_\mathrm{Li}$, and the electric current density $\Jflux$, read
\begin{gather}
    \label{eq:LiFluxElyte}
    \Nflux_\mathrm{Li} = \frac{\transf_\mathrm{Li}}{\Faraday} \frac{\mass_\mathrm{an}}{\mass_\mathrm{LiSalt}}  \Jflux - \porosity^\Brugge \diffcoeff^\mathrm{e}_\mathrm{Li} \frac{\partial}{\partial \xcoord} \Tilde{\redchempot}_\mathrm{Li},
    \\
    \label{eq:ElCurrDensElyte}
    \Jflux = - \porosity^\Brugge \conduc \frac{\partial}{\partial \xcoord} \altelpot - \porosity^\Brugge \conduc \frac{\transf_\mathrm{Li}}{\Faraday} \frac{\mass_\mathrm{an}}{\mass_\mathrm{LiSalt}}\frac{\partial}{\partial \xcoord} \Tilde{\redchempot}_\mathrm{Li}.
\end{gather}
Here, $\transf_\mathrm{Li}$, $\diffcoeff_\mathrm{Li}^\mathrm{e}$ and $\conduc$ denote the transference number of the Li-ions, the diffusion coefficient of the Li-ions and the ionic conductivity. Together they constitute the set of independent transport parameters of the electrolyte. Furthermore, $\Tilde{\redchempot}_\mathrm{Li}$ is the effective chemical potential of the Li-ions, and $\altelpot = \elpot + \redchempot_\mathrm{cat} \mass_\mathrm{an} / \Faraday \mass_\mathrm{IL}$ is the chemo-electric potential,\cite{Newman2012,Latz2015} defined with respect to the electrostatic Maxwell or Galvani potential $\elpot$, and the effective chemical potential $\tilde{\chempot}_\mathrm{cat}$ of the IL-cations. See \cref*{sec:SI_IL_Elyte_Theory} in the ESI and Ref.~\citenum{Schammer2020theory} for a detailed discussion of all relevant quantities.

\begin{table*}[!tb]
\begin{threeparttable}[b]
\centering
\begin{tabular}{rlcccc} 
    \toprule
    Parameter & Unit & Anode & Separator & Cathode & Electrolyte \\
    \midrule
    length $\Length^\mathrm{s,e}$ & \unit{\um} & 5 & 20 & 20 & 45 ($\Length_\mathrm{tot}$) \\
    (initial) radius $\Radius^\mathrm{s}_0$ & \unit{\um} & 0.15 & - & 5.5 [\citenum{Danner2016}] & - \\
    porosity $\porosity^\mathrm{s}$ & - & 0.781 \tnote{\dag} & 0.5 & 0.383 [\citenum{Danner2016}] & - \\
    spec. surface area $\Area_\mathrm{spec}^\mathrm{s}$ & \unit{\per \m} & \num{2.92e6} \tnote{\dag} & - & \num{3.37e5} \tnote{\dag} & - \\
    max. concentration $\conc_\mathrm{Li}^\mathrm{max,s}$ & \unit{\mol \per \cubic \m} & 311475 [\citenum{Verma2019}] & - & 36224 [\citenum{Danner2016}] & - \\
    rate constant $\RateConst^\mathrm{s}$ & \unit{\mol \per \m \squared \per \s} &  \num{2.06e-7} \tnote{\dag} & - & \num{7.51e-5} \tnote{\dag} & - \\
    diffusion coefficient $\diffcoeff_\mathrm{Li}^\mathrm{s,e}$ & \unit{\m \squared \per \s} & \num{e-18} [\citenum{Verma2019}] & - & \num{2e-15} [\citenum{Danner2016}] & \num{7.5e-12} [\citenum{Lorenz2023}] \\
    conductivity $\conduc$ & \unit{\siemens \per \m} & - & - & - & 0.192 [\citenum{Lorenz2022}] \\
    transference number $\transf_\mathrm{Li}$ & - & - & - & - & 0.1 [\citenum{Lorenz2022}] \\
    initial concentration $\conc_\mathrm{Li,0}^\mathrm{e}$ & \unit{\mol \per \cubic \m} & - & - & - & 2240 \tnote{\dag} \\
    \bottomrule
\end{tabular}
\begin{tablenotes}
    \item [\dag] calculated
\end{tablenotes}
\end{threeparttable}
\caption{Baseline cell parameters.}
\label{tab:Parameters}
\end{table*}

On the anode side, we model transport of the Li-ions in the Si NWs. Upon (de-) lithiation these undergo large volumetric changes. We use the chemo-mechanically coupled model from Kolzenberg \emph{et al.} to describe the deformation of the NWs via chemical expansion $\DeformGrad_\mathrm{chem}$ due to solid diffusion and mechanical deformations $\DeformGrad_\mathrm{elas}$, so that $\DeformGrad_\mathrm{rev} = \DeformGrad_\mathrm{elas} \DeformGrad_\mathrm{chem}$ (\cref*{eq:SI_DeformGrad} in the ESI).\cite{Castelli2021,Kolzenberg2022a} The concentration gradients that build up during (de-)lithiation cause mechanical deformations and, thus, mechanical stresses in the NWs. \Cref{tab:SystemOfEquations} comprises the corresponding two differential equations expressed in cylindrical coordinates. The two equations describe the mutually coupled solid diffusion affecting the time evolution of the concentration and the reversible stresses which obey the momentum balance (see \cref*{eq:SI_revDeform,eq:SI_chemPotStress}). Note that the evolution of the Si NWs is expressed using a Lagrangian description which is fixed to the material points of the NWs at a given reference configuration (here the undeformed initial state at time $t=0$, denoted by a subscript "0"). Here, $\Radius_0^\mathrm{a}$ is the initial NW radius, $\FirstPiola_\mathrm{rev}$ is the first reversible Piola-Kirchhoff stress tensor, $\diffcoeff^\mathrm{a}_\mathrm{Li}$ the solid diffusion coefficient of Li in Si, and $\soc^\mathrm{a}_\mathrm{Li} = \conc^\mathrm{a}_\mathrm{Li,0}/\conc_\mathrm{Li,0}^\mathrm{a,max}$ is the dimensionless concentration or state of charge (SoC) with $\conc_\mathrm{Li,0}^\mathrm{a,max}$ being the maximum possible concentration of Li in Si. For a more detailed discussion, we refer to \cref*{sec:SI_Si_Anode_Theory} in the ESI and Refs.~\citenum{Castelli2021,Kolzenberg2022a}.

On the cathode side, we use a simple model for solid diffusion based on Fick's law of diffusion. For simplicity, we assume spherical active particles in the cathode and express the corresponding transport equation via spherical coordinates, see \cref{tab:SystemOfEquations}. Here, $\Radius^\mathrm{c}$ is the radius of the spherical particles, $\diffcoeff^\mathrm{c}_\mathrm{Li}$ the solid diffusion coefficient of Li in the cathode material and $\soc^\mathrm{c}_\mathrm{Li}$ is the SoC of the cathode.

Finally, Li-ion transport in the solid and liquid phases is coupled via source terms based on a standard Butler-Volmer approach,\cite{Latz2013}
\begin{equation}
    \source_\mathrm{Li} = \Area^\mathrm{s}_\mathrm{spec} \ExCurrDens^\mathrm{se}/\Faraday,
\end{equation}
where $\Area^\mathrm{s}_\mathrm{spec}$ is the specific surface area of the respective electrode. The electrode-electrolyte current density $\ExCurrDens^\mathrm{se}$ reads
\begin{equation}
    \ExCurrDens^\mathrm{se} = 2 \ExCurrDens_0^\mathrm{se} \sinh \left( \frac{\Faraday}{2 \Rconst \Temp} \OverPot^\mathrm{se} \right),
\end{equation}
and depends on the overpotential $\OverPot^\mathrm{se}$ and the exchange current density $\ExCurrDens_0^\mathrm{se}$. Here, 
\begin{equation}
    \ExCurrDens_0^\mathrm{se} = \Faraday \RateConst^\mathrm{s} \left( \soc_\mathrm{Li}^\mathrm{s}(1-\soc_\mathrm{Li}^\mathrm{s}) \right)^{0.5} \left( \frac{\conc_\mathrm{Li}^\mathrm{e}}{\conc_\mathrm{Li,0}^\mathrm{e}} \right)^{0.5}.
\end{equation}
where $\RateConst^\mathrm{s}$ is the rate constant, $\soc_\mathrm{Li}^\mathrm{s}$ the SoC of the electrode and $\conc_\mathrm{Li,0}^\mathrm{e}$ the initial Li concentration of the electrolyte.
The overpotential is defined as
\begin{equation}
    \OverPot^\mathrm{se} = \elpot^\mathrm{s} - \elpot^\mathrm{e} - \voltage^\mathrm{s}_0(\soc^\mathrm{s}_\mathrm{Li}) - \frac{\Rconst \Temp}{\Faraday} \ln\left( \frac{\conc_\mathrm{Li}^\mathrm{e}}{\conc_\mathrm{Li,0}^\mathrm{e}} \right),
\end{equation}
where $\elpot^\mathrm{s}$ and $\elpot^\mathrm{e}$ are the electric potential of the electrode and electrolyte, respectively. $\voltage^\mathrm{s}_0(\soc^\mathrm{s}_\mathrm{Li})$ is the half-cell open circuit potential at the respective SoC. The externally applied current density couples to the electrodes via $\Jflux_\mathrm{ext} = \Faraday \source_\mathrm{Li} \mathrm{d}\xcoord$.

We use the following boundary conditions for the differential equations in \cref{tab:SystemOfEquations}. The flux density $\ExCurrDens^\mathrm{se}/\Faraday$ is the boundary condition for the anode and cathode diffusion equations at the particle/NW surface ($\rcoord = \Radius^\mathrm{s}$). At the center ($\rcoord = 0$), no flux is assumed due to the radial symmetry of the particle/NW ($\partial \conc_\mathrm{Li}^\mathrm{s} / \partial \rcoord = 0$). In the stress equation, the radial component of the Piola stress vanishes ($\FirstPiola_\mathrm{rev,r} = 0$) at the NW surface ($\rcoord = \Radius^\mathrm{a}$) because we assume that the NWs can expand freely into the electrolyte as long as there is enough pore space left. Assuming a primitive cubic lattice of NWs, we show in \cref*{sec:SI_MinPoro} in the ESI that the porosity threshold for the anode is $\porosity^\mathrm{a} = 22\%$ for geometry reasons. For the electrolyte transport equations, we assume that no flux can enter the current collectors at $\xcoord = 0$ (anode side) and $\xcoord = \Length_\mathrm{tot}$ (cathode side). Thus, there, we set $\partial \conc_\mathrm{Li}^\mathrm{e} / \partial \xcoord = \partial \elpot / \partial \xcoord = 0$. Also, $\vel = 0$ at $\xcoord = 0$ with an open boundary condition at $\xcoord = \Length_\mathrm{tot}$ for numerical reasons.

\subsection{Parameterization}
\label{sec:Parameterization}

In the following, we summarize the parameters that we use for our baseline simulations. For the baseline cell, we parameterize the amorphous Si NW anode similar to the experimental works of Ryan \emph{et al.},\cite{Kim2017,Stokes2019_2,Stokes2020,Karimi2023} use a standard NMC111 cathode, and an ionic liquid electrolyte of the form IL + Li-salt with common anion. Here, we use the mixture (0.4)LiFSI(0.6)Pyr12O1FSI (see \cref*{sec:SI_ILElyte} in the ESI for chemical acronyms) which has a reasonably high Li-ion transference number of about 0.1.\cite{Lorenz2022,Kilchert2023,Lorenz2023} The partial molar volumes of the three ionic species are $\pmv_\mathrm{FSI} = $ \qty{8.92e-5}{\cubic \m \per \mol}, $\pmv_\mathrm{Pyr} = $ \qty{1.48e-4}{\cubic \m \per \mol} and $\pmv_\mathrm{Li} = $ \qty{1.11e-6}{\cubic \m \per \mol}.\cite{Lorenz2022} 

It has been reported that Si nanostructures are less prone to fracture if they have a diameter below roughly \qty{300}{\nm}.\cite{Ryu2011,Liu2012a} Thus, for our baseline cell, we assume a cylindrical geometry with $\Radius^\mathrm{a}_0 =$ \qty{150}{\nm} radius and a length of $\Length^\mathrm{a} =$ \qty{5}{\um} for our Si NWs. For the anode, we assume an active mass loading of \qty{0.25}{\mg \per \cm \squared}, which results in a porosity of 0.78 (see \cref*{sec:SI_Poro_ActiveLoad}). Note that the influence of these values on the cell performance will be investigated in \cref{sec:Param_Study}. For the anode open circuit voltage (OCV) curve $\voltage^\mathrm{a}_0(\soc^\mathrm{a}_\mathrm{Li})$, we use the GITT data from Ref.~\citenum{Pan2019} (see \cref*{eq:SI_OCVano} in \cref*{sec:SI_OCV}). For the cathode, the OCV curve $\voltage^\mathrm{c}_0(\soc^\mathrm{c}_\mathrm{Li})$ is taken from Ref.~\citenum{Danner2016}, \emph{cf.} \cref*{eq:SI_OCVcat}. To make sure that the Si anode is limiting cell performance, we set its capacity slightly lower than the cathode capacity and choose the cathode thickness (length) accordingly. This ensures that the cathode can supply enough Li to the anode. Specific surface areas are derived from the respective electrode geometries (length, radius and porosity), see \cref*{eq:SI_SpecAreaCat,eq:SI_SpecAreaAno} in \cref*{sec:SI_SpecArea}. In our simulations, we terminate the process of charging (/discharging) the cell once the NW surface (outermost voxel in radial direction) reaches a SoC of 0.95 (/0.05). Thereby, we avoid the onset of Li plating, which typically occurs for Li-saturated electrode particles.\cite{Waldmann2018} All simulations are performed at room temperature ($\Temp = \qty{298.15}{\kelvin}$). \Cref{tab:Parameters} lists the most important parameters.

\section{Simulation Results and Discussion}
\label{sec:Results}

In this section, we present numerical results obtained from computer simulations performed with our cell model. First, in \cref{sec:Baseline_Sim}, we discuss our "baseline" system with the parameterization presented in \cref{sec:Parameterization}. Here, we focus on the state of charge (SoC) of the electrodes and on the stresses in the Si nanowires (NWs) which are hardly accessible via experiments. Next, we discuss the influence of the silicon geometry and material parameters on the cell-performance. For this purpose, in \cref{sec:Param_Study}, we perform parameter studies focusing on the Si NW anode. In \cref{sec:VaryDsRano}, we investigate the diffusion of Li inside the Si NWs by varying the solid diffusion coefficient and the NW radius while maintaining a constant anode capacity. Finally, in \cref{sec:VaryPoros}, we investigate the effect of anode porosity on the capacity of the cell.

\subsection{Baseline Simulation}
\label{sec:Baseline_Sim}

In this section, we perform full cell simulations of the as above parameterized battery model, which consist of one full charge-discharge cycle. 

\begin{figure}[tb]
	\centering
	\includegraphics[width=0.45\textwidth]{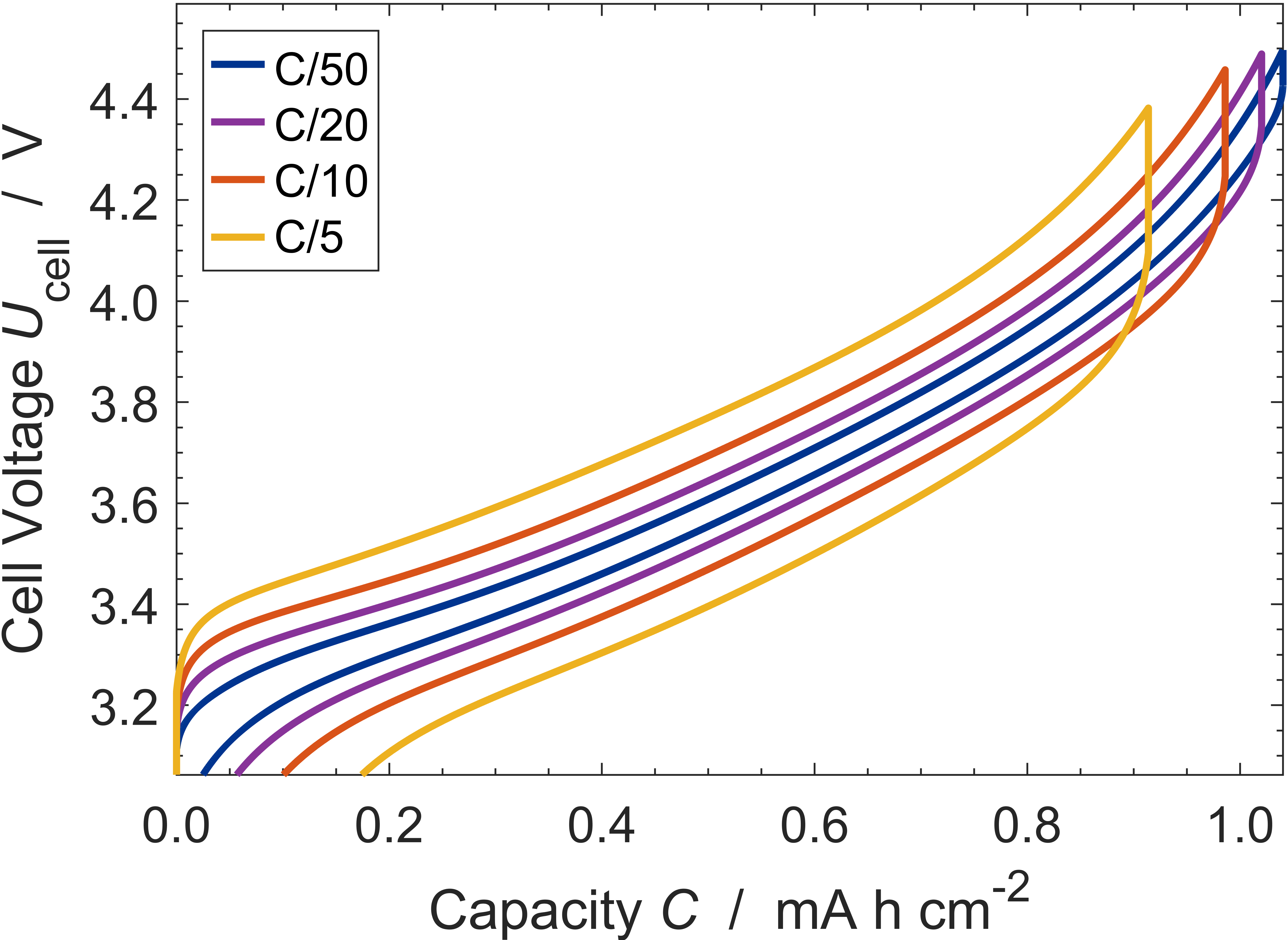}
	\caption{Cell potential for one charge-discharge cycle versus areal capacity for different C-rates.}
	\label{fig:CellVolt}
\end{figure}

\Cref{fig:CellVolt} shows typical charge-discharge curves of the baseline cell, where the cell potential is plotted against the areal capacity at four different C-rates ranging from C/50 to C/5. For all C-rates, charging starts at \qty{3.06}{\volt} and is stopped once the SoC of the Si NWs reaches the above mentioned value of 0.95. Apparently, the shape of the profiles is similar for all C-rates. 
However, increasing the charging dynamics (higher C-rates) reduces the terminal voltage at end of charge (EoC) as well as the maximally achievable capacity. In our simulation, end of discharge (EoD) is reached once the SoC of the anode is 0.05. It can be seen for all C-rates, that the final capacity does not reach zero again. This deviation increases with higher C-rates. We go into more detail on reasons for this observation below in this section.

\begin{figure}[tb]
	\centering
	\includegraphics[width=0.45\textwidth]{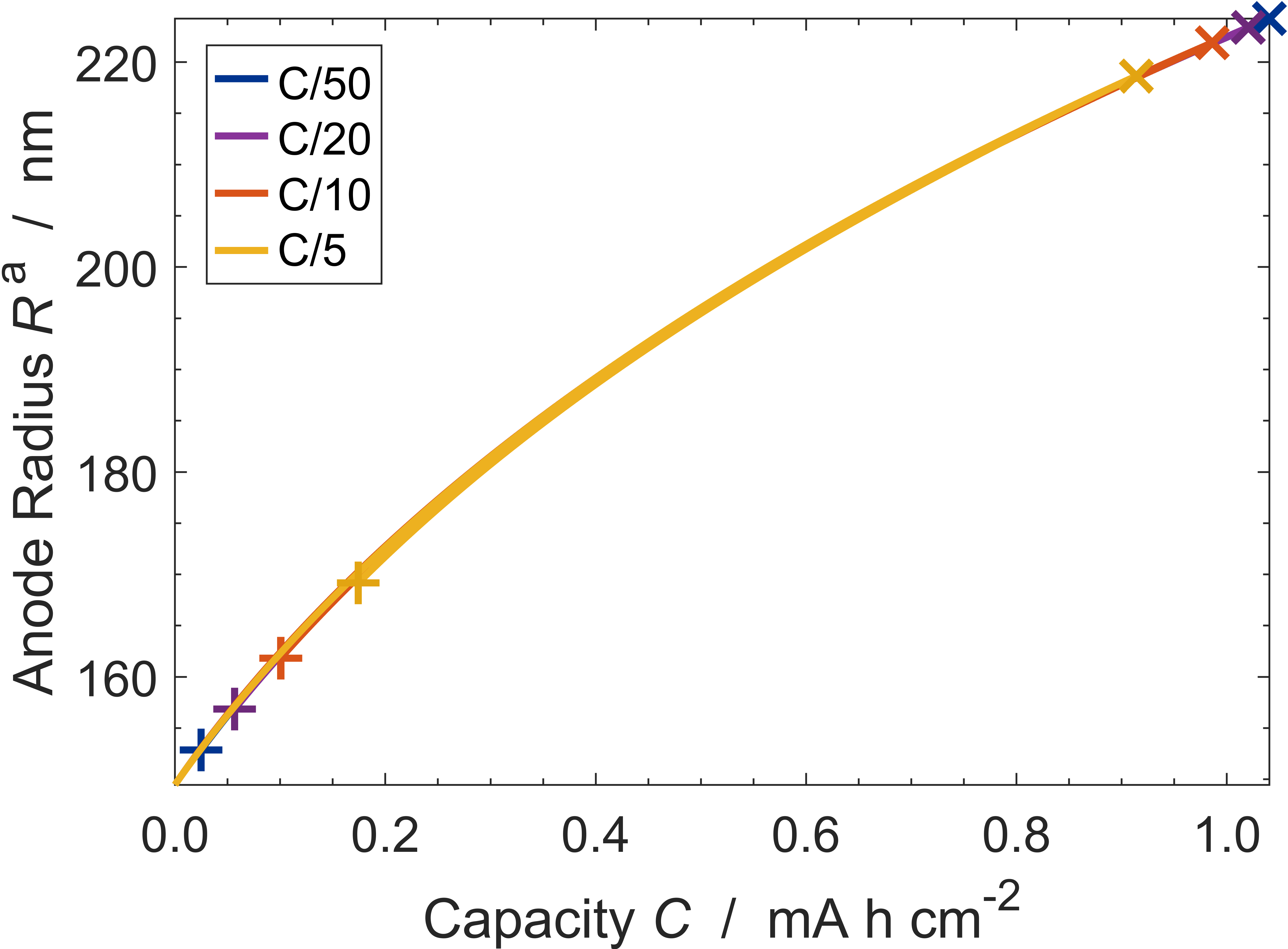}
	\caption{Evolution of the Si NW radius over one charge-discharge cycle for different C-rates. Symbols mark the end of charge (x) and the end of discharge (+).}
	\label{fig:Rano}
\end{figure}

\begin{figure*}[tb]
	\centering
	\includegraphics[width=0.85\textwidth]{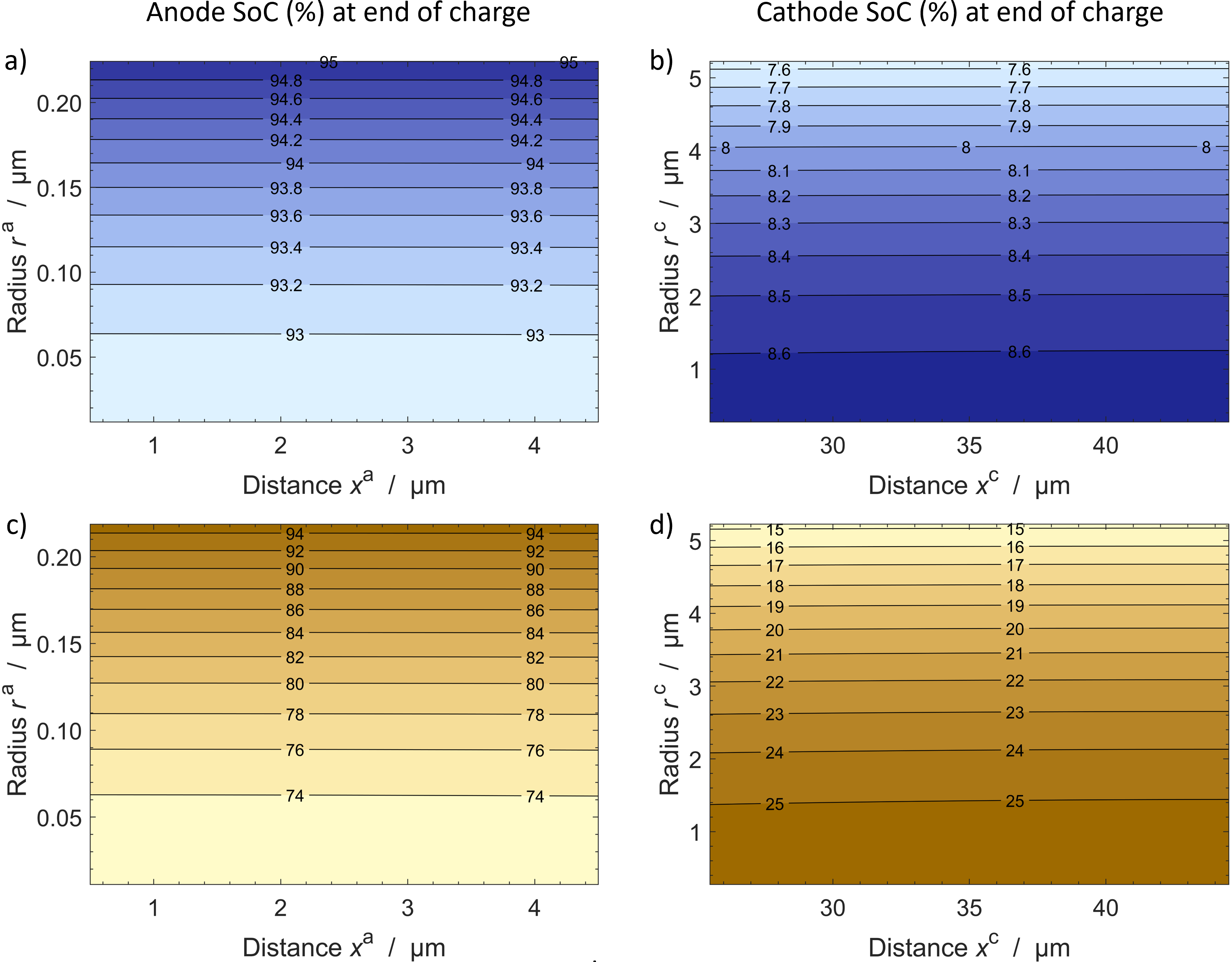}
	\caption{Spatial distribution of the state of charge (SoC in \%) in the anode (a) and c)) and the cathode (b) and d)) at the end of charge after charging with C/50 (blue, a), b)) and C/5 (yellow, c), d)).}
	\label{fig:SOCanocat}
\end{figure*}

Next, we focus on the reversible deformations of the Si NWs. \Cref{fig:Rano} shows the evolution of the NW radius $\Radius^\mathrm{a}$ as function of the capacity during one charge-discharge cycle, evaluated at four different C-rates. At the beginning of charging, the NWs are in their pristine state with radius $\Radius^\mathrm{a}_0 = \qty{150}{\nm}$. Apparently, during charging, the radius increases up to a terminal value at the end of charge (marked with 'x'). This can be attributed to the fact that during charging, Li diffuses into the NWs, which results in a radial "swelling". The degree of swelling, \emph{i.e.} the maximal radius at EoC, decreases with increasing charge rate. Once discharging starts, the radii decrease along the path of charging until the end of discharge (marked with '+'). This results in an overlap of the curves for different C-rates indicating mostly chemical and only small elastic deformation. However, it can be seen that the terminal radii at EoD are larger than the initial radius. Thus, the NWs do not fully contract back to the initial configuration and the swelling is not fully reversed. Apparently, the residual swelling at EoD increases with increasing C-rates. This is in agreement with the observation from \cref{fig:CellVolt}.

Next, we investigate the SoC-profile of Li within the solid, active materials of the electrodes at two different C-rates at the end of charge. The SoC is a spatially resolved quantity along the two dimensions of our model, see also \cref{fig:CellScheme}, namely the radial coordinate $\rcoord$ of the electrode particles/NWs and the lateral extension $\xcoord$ of the cell. \Cref{fig:SOCanocat} shows four contour plots for the Li content (in percent) versus these two spatial dimensions. \Cref{fig:SOCanocat}~a) and c) show the SoC of the anode at the EoC for a C-rate of C/50 (blue, a)) and C/5 (yellow, c)). Similarly, \cref{fig:SOCanocat}~b) and d) show the SoC of the cathode at the EoC with the respective C-rates C/50 (blue, b)) and C/5 (yellow, d)). As can be inferred from figures b) and d), at the EoC (anode SoC of 95\%), there is still capacity left in the cathode (>5\%) for both C-rates. This property of the electrodes is an intended artifact of our parameterization protocol, where we set the cathode capacity to be slightly larger than the anode capacity (see \cref{sec:Parameterization}). Apparently, in each plot, the SoC is almost constant along the $\xcoord$-dimension. This suggests that the transport of Li-ions in the IL electrolyte is fast enough as to sustain a sufficient Li supply everywhere in the porous electrodes. Along the radial dimension (from the center to the outermost voxel), however, there is a small SoC-gradient visible of roughly 1\% in the cathode (\cref{fig:SOCanocat}~b)) and 2\% in the anode (\cref{fig:SOCanocat}~a)) for the moderate C-rate of C/50. This changes when we go to a higher C-rate of C/5. \Cref{fig:SOCanocat}~c) shows that when the outermost region of the NWs reach their terminal SoC (95\%), the innermost regions are not yet fully charged. This leads to a significant gradient in the SoC-profile along the $\rcoord$-dimension. As shown by \cref{fig:SOCanocat}~d), the increased C-rate also amplifies the radial SoC gradient in the cathode particles. Altogether, this suggests that for higher applied currents the transport of Li becomes diffusion-limited inside the Si NWs. Hence, for higher C-rates, the fraction of the available anode capacity, which is provided by the inner regions of the NWs, cannot be fully exploited. This explains the decline in cell capacity under enhanced C-rates as shown in \cref{fig:CellVolt} and the effect that enhanced charging currents lead to reduced swelling of the NWs (see \cref{fig:Rano}).

\begin{figure}[tb]
	\centering
	\includegraphics[width=0.45\textwidth]{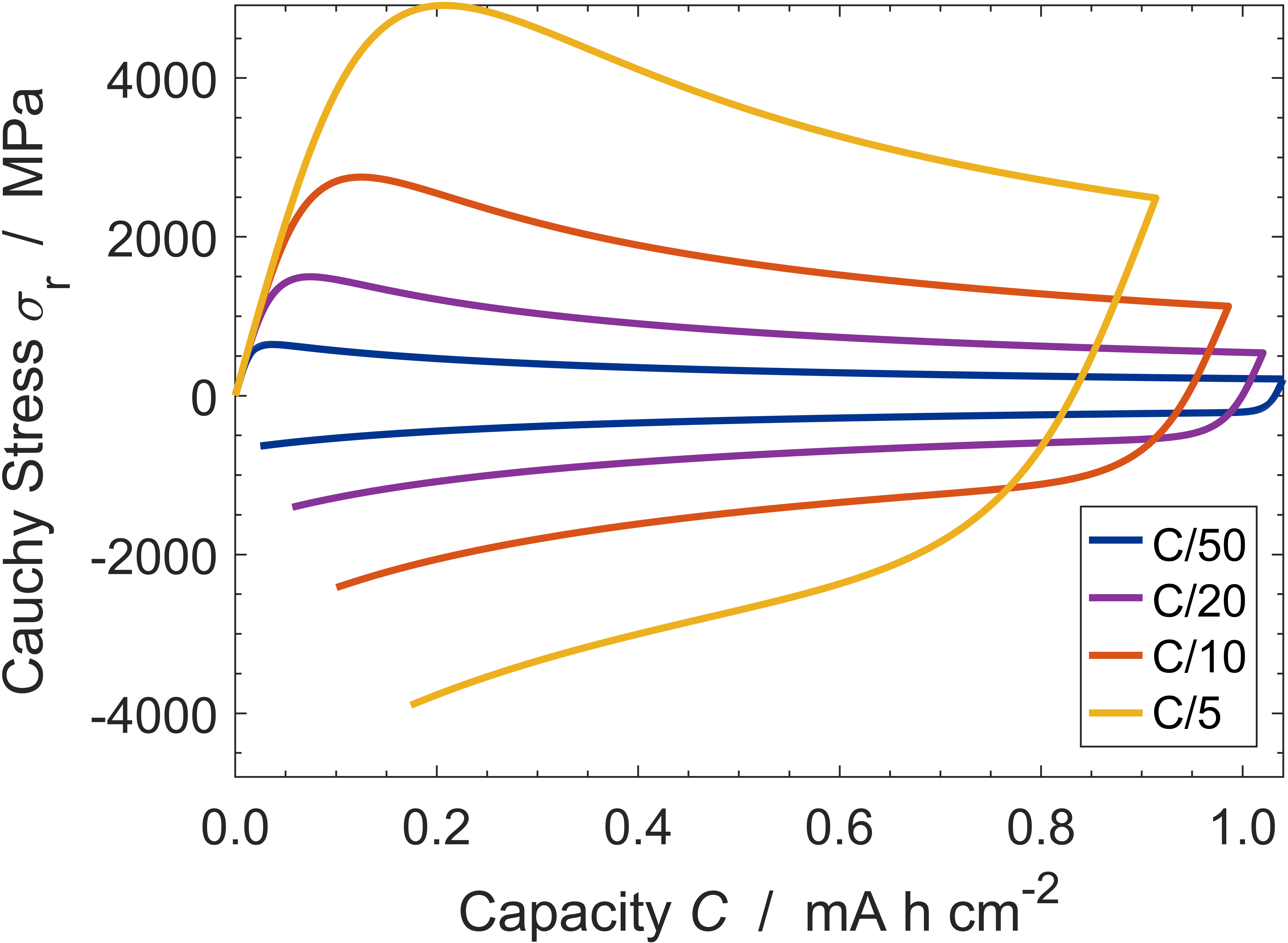}
	\caption{Evolution of the radial component of the Cauchy stress in the center of the Si NWs over one charge-discharge cycle for different C-rates.}
	\label{fig:CauchyStressRvarCurr}
\end{figure}

\begin{figure}[tb]
	\centering
	\includegraphics[width=0.45\textwidth]{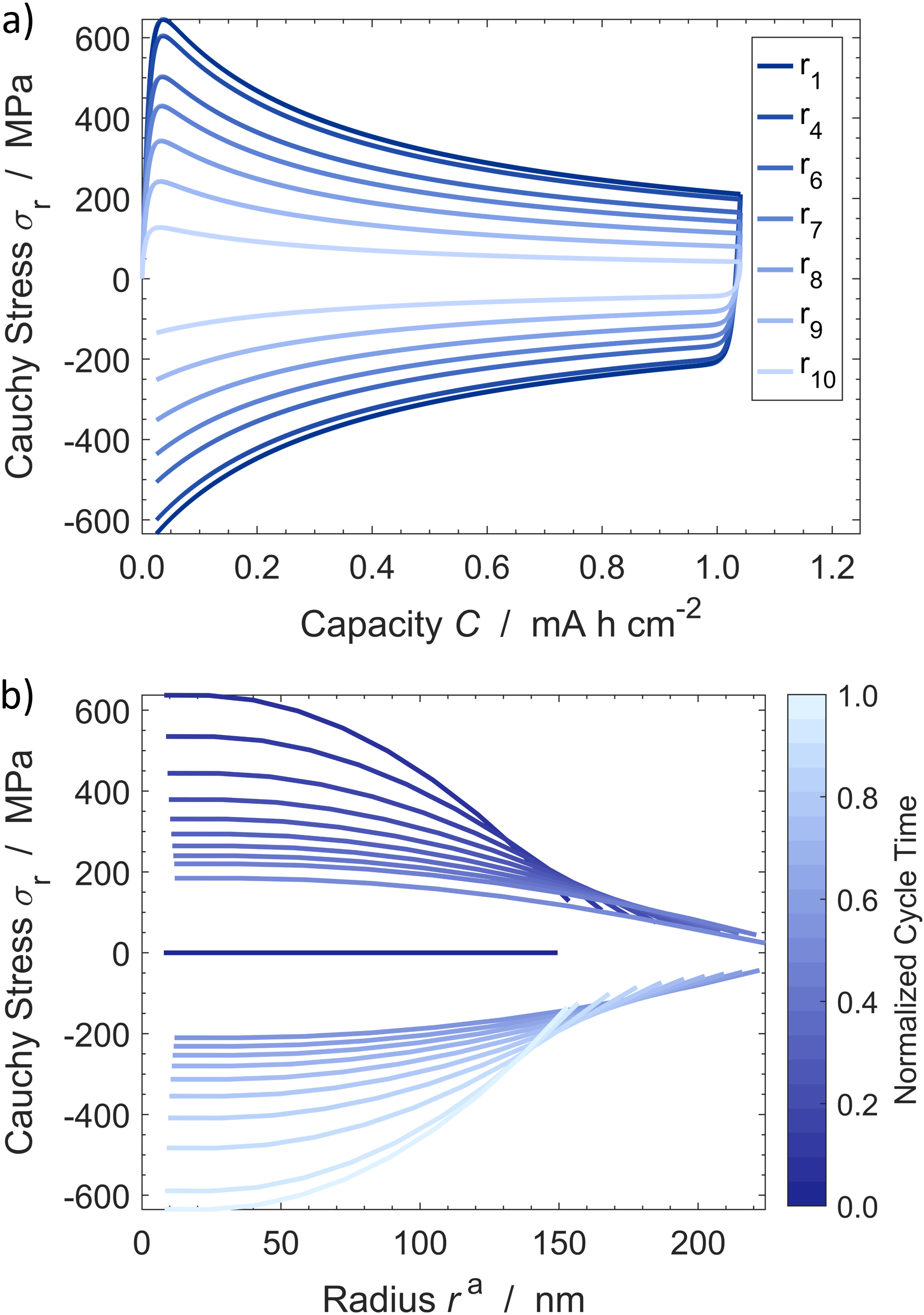}
	\caption{Evolution of the radial component of the Cauchy stress within the NW shown a) versus cell capacity for different radial positions within the NW (from $\rcoord_1$ in the center to $\rcoord_{10}$ near the edge) and b) versus the NW radius for various times (corresponding to different anode SoCs) during one full cycle.}
	\label{fig:CauchyStressR1}
\end{figure}

Finally, we investigate the mechanical strain which is induced upon the NWs during (de-)lithiation. For this purpose, we focus on the evolution of the radial components $\Cauchy_\mathrm{r}$ of the Cauchy stress $\Cauchy$ inside the NWs, as given by \cref*{eq:SI_CauchyFirstPiola} (in the ESI, we also discuss the tangential component of the Cauchy stress $\Cauchy_\upphi$, see \cref*{fig:SI_CauchyStressPhiRvarCurr,fig:SI_CauchyStressPhi1} in \cref*{sec:SI_TangentialCauchy}). In the ESI, we show that there exist no significant stress gradients along the $\xcoord$-coordinate (see \cref*{fig:SI_StressRvsX} in \cref*{sec:SI_CauchyX} in the ESI) which is in line with the absence of a concentration gradient along the $\xcoord$-dimension (\emph{cf.} \cref{fig:SOCanocat}). Hence, the position of the NW along the $\xcoord$-direction has a negligible influence on the stress distribution within the NW and it suffices to study the NW stresses at one particular location. In the following, we restrict our discussion of the NW stresses to the $\xcoord$-voxel closest to the separator. 

\Cref{fig:CauchyStressRvarCurr} shows the radial component of the Cauchy stress over the capacity for different C-rates at the center of the NW (voxel '$\rcoord_1$'). The Cauchy stress is calculated relative to the present radius of the NWs and, thus, defined in the Euler frame of reference. Apparently, the stress profile has a similar shape for all C-rates, with tensile (positive) stresses during charging which soon become compressive (negative) stresses during discharging. For higher C-rates the absolute stress increases. The initial steep increase of the stress, can be attributed to the large concentration gradients that build up when the NW surface gets lithiated. Tensile stresses stretch the inner part of the NWs to accommodate the chemical expansion of the outer region. With further lithiation the NW radially expands (\emph{cf.} \cref{fig:Rano}). Thus, the concentration difference spreads over an increasing radius causing the stress to slowly decrease. The same effect happens inversely during discharging, where the surface starts to get delithiated compressing the NW and the radius decreases. The discharge process terminates when there is still some capacity left in the NWs. Similarly to \cref{fig:SOCanocat}~c) and d) which show the SOC of the electrodes at the end of charging (EoC), there remains a concentration gradient in the anode at the end of discharge (EoD). Thus, the stress is not fully released again.

Next, we focus on the evolution of the radial stresses at the lowest C-rate (C/50, blue), as shown in \cref{fig:CauchyStressR1}. In order to study the stress evolution at different radial distances $\rcoord_\mathrm{m}$ from the NW center, we designate $\mathrm{m} = 10$ locations $(\rcoord_1,\ldots,\rcoord_{10})$ inside the NW with increasing distance from the center of the NW (where $\rcoord_1$ is near the center of the NW and $\rcoord_{10}$ is near the edge). However, because the radii swell and shrink during cell operation, the magnitudes of the distances $\rcoord_\mathrm{m}$ are time dependent (depend on the capacity / SoC). \Cref{fig:CauchyStressR1}~a) depicts the stress versus capacity at these positions. Here, the largest curve at voxel $\rcoord_1$ equals the blue curve in \cref{fig:CauchyStressRvarCurr}. Apparently, the absolute values of the stresses decrease with increasing radial distance from the NW center. This observation is consistent with the boundary condition of vanishing radial stresses at the edge of the NW. Note that we assume that the NWs can expand freely, as long as there is enough pore space left. In \cref{fig:CauchyStressR1}~b) the plotting dimensions are interchanged and the stress is now plotted against the radial dimension of the NW for various cycle times. It can be seen that the maximal radius increases during charging and decreases during discharging (in accordance with \cref{fig:Rano}), which stems from the swelling property of the NWs. Apparently, at any given time during the cycle, the absolute stresses decrease with increasing radial dimension, which is consistent with the behaviour shown in \cref{fig:CauchyStressR1}~a). The dark, flat curve at $\Cauchy_\mathrm{r} = 0$ marks the start of the simulation. After a steep increase in the beginning -- as also visible in \cref{fig:CauchyStressR1}~a) -- the tensile stress slowly decreases during charging and changes to compressive stress during discharging. In the ESI, we present the same discussion for the tangential component of the Cauchy stress $\Cauchy_\upphi$, see \cref*{fig:SI_CauchyStressPhiRvarCurr,fig:SI_CauchyStressPhi1}. Altogether, the general behaviour of the stress evolution is in very good agreement with the results for spherical Si particles presented in Ref.~\citenum{Kolzenberg2022a} (see, especially, Fig.~7 there).

\subsection{Parameter Studies}
\label{sec:Param_Study}

In this section, we investigate the influence of some key parameters of the anode on the overall cell performance. First, in \cref{sec:VaryDsRano}, we study the cumulative influence of the solid diffusion coefficient and the NW radius on the cell capacity. Second, in \cref{sec:VaryPoros}, we focus on the influence of the anode porosity, and how it affects the maximal cell capacity.

\subsubsection{Influence of NW Radius and Solid Diffusion Coefficient}
\label{sec:VaryDsRano}

In \cref{sec:Baseline_Sim}, we concluded from our simulations of the baseline cell that the cell performance can be negatively influenced by diffusion limitations inside the Si NWs. In this section, we aim for a more detailed understanding of this effect. To address this goal, we focus on the solid diffusion coefficient $\diffcoeff^\mathrm{a}_\mathrm{Li}$ of Li in the Si NWs.

A wide range of diffusion coefficients for Li in Si, spanning several orders of magnitude, has been reported in the literature.\cite{Soni2012,Wang2016,Pan2019,Verma2019} For example, Verma \emph{et al.} report values ranging from \qty{1e-18}{\m \squared \per \s} to \qty{1e-16}{\m \squared \per \s} for nanostructured Si, and propose the value \qty{1e-18}{\m \squared \per \s} for NW-like structures (this value was also used for the parameterization of the baseline cell discussed in \cref{sec:Baseline_Sim}, see also \cref{tab:Parameters} in \cref{sec:Parameterization}).\cite{Verma2019} In our parameter study, we vary both the solid diffusion coefficient $\diffcoeff^\mathrm{a}_\mathrm{Li}$, and the initial radius of the Si NWs $\Radius^\mathrm{a}_0$. As we have found in \cref{sec:Baseline_Sim}, both parameters influence how well and uniformly the anode is lithiated during charging. We emphasize that the initial anode porosity is constant during this parameter study. Increasing the NW radius implies less but thicker NWs per area, whereas the pore space and, thus, distance between the NWs is unaffected. In particular, varying the radius of the NWs does not change the active mass loading of the anode (see also \cref*{eq:SI_ActiveLoad} in \cref*{sec:SI_Poro_ActiveLoad} in the ESI). 

\begin{figure}[tb]
	\centering
	\includegraphics[width=0.48\textwidth]{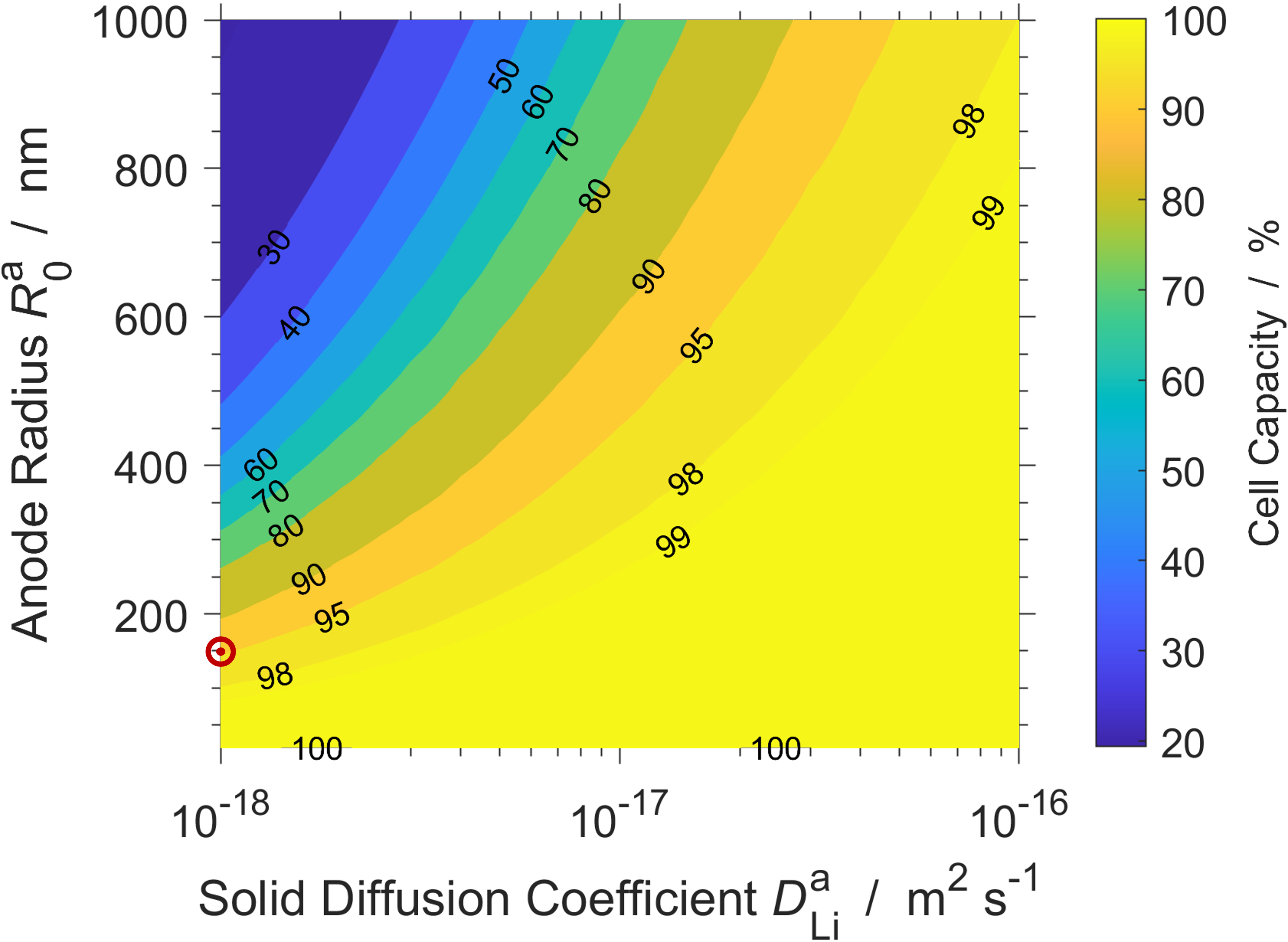}
	\caption{Normalized cell capacity in \% at the end of charge after charging with a C/10-rate versus Si NW radius and anode solid diffusion coefficient. The red circle marks the parameter set of the baseline cell.}
	\label{fig:Cap_vs_D_R}
\end{figure}

Our parameter set for $\diffcoeff^\mathrm{a}_\mathrm{Li}$ comprises a total of 27 values, ranging from \qty{1e-18}{\m \squared \per \s} up to \qty{1e-16}{\m \squared \per \s}, as reported in the literature.\cite{Verma2019} In addition, the NW radius is varied from \qty{20}{\nm} up to \qty{1}{\um} in steps of \qty{20}{\nm} (50 values). For each pair of these two parameters we perform a full cell simulation with a C/10-rate, and extract the maximal cell capacity at the end of charge. This results in a database of $27 \times 50 = 1350$ capacity values. \Cref{fig:Cap_vs_D_R} illustrates the result database as a contour plot for the capacity values versus the input parameters shown on the $\xcoord$-axis (diffusion coefficients) and on the $y$-axis (anode radius). The capacity is normalized with respect to the largest capacity value obtained. Because the values for the diffusion coefficient span two orders of magnitude, the $\xcoord$-axis is set to a logarithmic scale. The anode radius on the $y$-axis displays the initial radius of the pristine NWs (at beginning of simulation). The red circle marks the value-set for the two parameters that were used for the baseline simulation.

Apparently, for lower diffusion coefficients in the order of \qty{1e-18}{\m \squared \per \s}, the maximal cell capacity decreases significantly with increasing NW radius. This property can be attributed to the slow Li diffusion inside the Si NWs and was also observed in the baseline simulations from the last section for different C-rates (see  \cref{fig:CellVolt,fig:SOCanocat}). For a higher solid diffusion coefficient in the order of \qty{1e-17}{\m \squared \per \s}, the NW radius can be increased up to about \qty{600}{\nm} without loosing more than 10\% of the maximal achievable capacity. Upon further increase of the solid diffusion coefficient, only minimal capacity losses are to be expected from using thicker NWs. We emphasize that these conclusions are based on the assumption that the electrolyte can supply enough Li to the anode. In our model, this is the case, and the electrolyte is not limiting the overall transport in the investigated parameter range.

Altogether, we conclude that nanostructured Si anodes with small diameter are more effective because their full capacity can be exploited during charging. For thicker structures, however, slow diffusion of Li inside the Si NWs limits the overall achievable cell capacity.

\subsubsection{Influence of Anode Porosity}
\label{sec:VaryPoros}

In the second parameter study, we investigate the influence of the initial anode porosity $\porosity^\mathrm{a}_0$ on the cell capacity. In \cref{sec:Model}, we defined the porosity of an electrode as the pore volume fraction of the respective electrode that contains the electrolyte. In the case of our Si NW anode, the porosity changes during cycling due to the expansion and contraction of the Si NWs. In our notation we, thus, distinguish between the initial porosity of the anode $\porosity^\mathrm{a}_0$ (pristine state) and the porosity at the end of charge $\porosity^\mathrm{a}_\mathrm{EoC}$.

Studying the effect of varying the initial anode porosity is relevant because of the large volumetric changes of Si. As shown above in \cref{fig:Rano}, the radius of the Si NWs can increase by up to 50\% of its initial value during lithiation. Thus, at some point the NWs can start to touch which creates large stresses in the material and aggravates the degradation of the anode. Therefore, in our simulations, the charging of the cell stops once the NWs have no more space to expand freely. This is reflected in a threshold for the minimal porosity which is solely determined by the underlying geometry of the anode. Assuming a primitive cubic lattice of NWs the minimal porosity equals 0.215 (see also \cref*{sec:SI_MinPoro} in the ESI).

For our study, we increase the cathode thickness to $\Length^\mathrm{c}=\qty{31}{\micro\meter}$ to always ensure a sufficient Li supply for the anode. We perform full cell simulations at a C/10-rate of the baseline cell described above (see \cref{sec:Parameterization}) and vary the pore space of the pristine anode $\porosity^\mathrm{a}_0$ from 0.5 to 0.8 in steps of 0.01 (31 values) while keeping the initial NW radius fixed.

\begin{figure}[tb]
	\centering
	\includegraphics[width=0.48\textwidth]{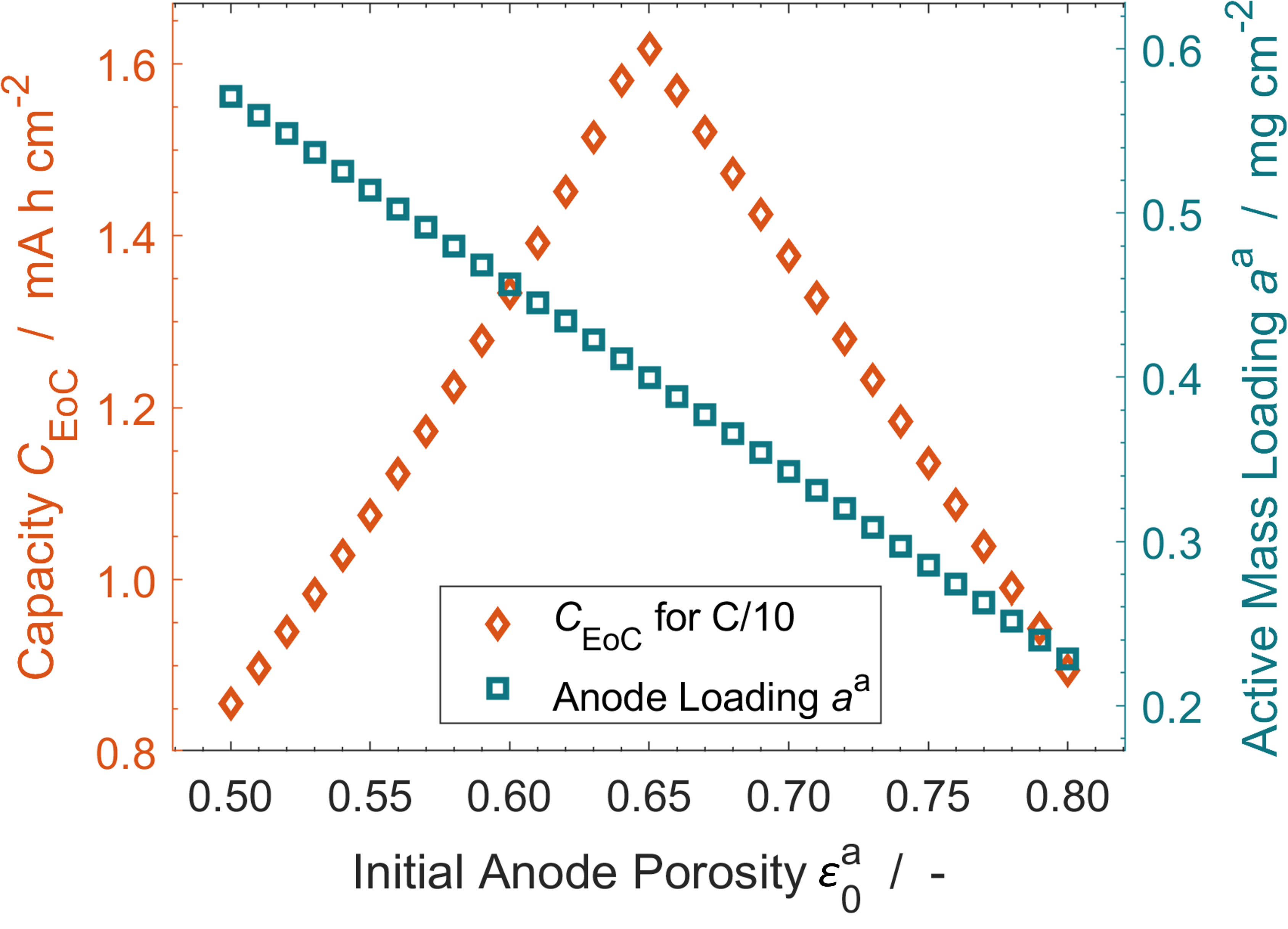}
	\caption{Maximum cell capacity at the end of charge after charging with a C/10-rate (diamonds, left axis) and anode active mass loading (squares, right axis) versus initial anode porosity.}
	\label{fig:Cap_vs_eps_ano}
\end{figure}

It is important to note that changing the initial porosity of the anode while keeping the initial radius of the NWs untouched is equivalent to a change in distance between the NWs. Thus, less pore space means more NWs per area (a more dense anode). This makes the initial anode porosity $\porosity^\mathrm{a}_0$ inversely proportional to the active mass loading of the anode $\Load^\mathrm{a}$. This relation is visualized in \cref{fig:Cap_vs_eps_ano} (squares).

\Cref{fig:Cap_vs_eps_ano} shows the maximum cell capacity at the end of charge $C_\mathrm{EoC}$ for the different initial anode porosities $\porosity^\mathrm{a}_0$ (diamonds). This capacity exhibits a maximum around 0.65 initial anode pore space ($C_\mathrm{EoC}^\mathrm{max} \approx \qty{1.62}{\milli\ampere \hour \per \centi\meter\squared}$). Upon further increasing the porosity, the capacity decreases. This decline is a direct consequence of the decreasing active mass loading of the anode. However, the capacity decline towards lower initial anode porosity (higher anode mass loading) is a consequence of the expansion of the Si NWs and the depletion of the anode pore space.

\begin{figure}[tb]
	\centering
	\includegraphics[width=0.47\textwidth]{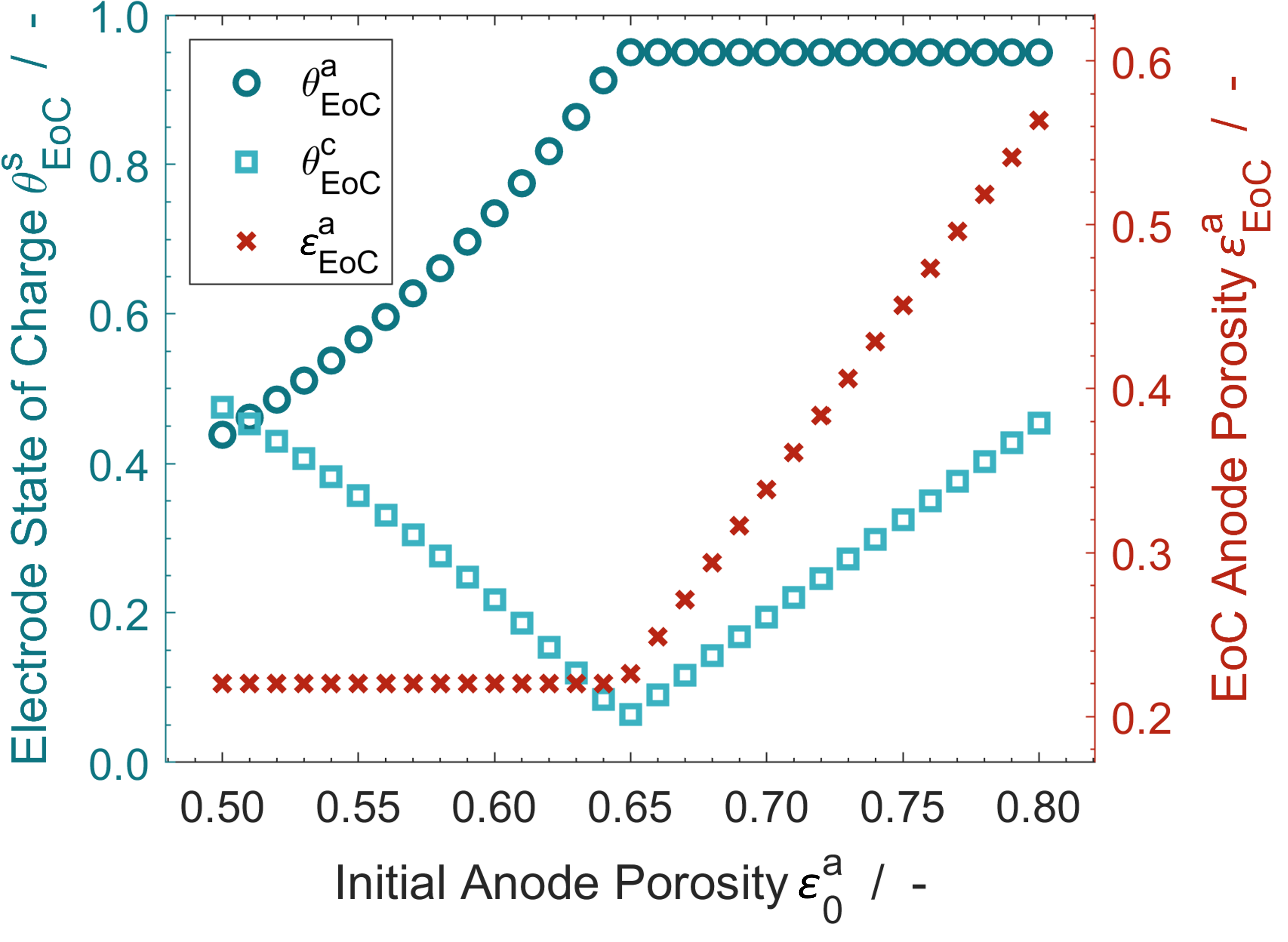}
	\caption{SoC (left axis) of anode (circles) and cathode (squares) and anode porosity (crosses, right axis) at the end of charge after charging with a C/10-rate versus initial anode porosity.}
	\label{fig:SOC_eps}
\end{figure}

The data shown in \cref{fig:SOC_eps} helps to obtain a better understanding of this effect. \Cref{fig:SOC_eps} visualizes the SoC of cathode and anode as well as the anode porosity at the end of charge at the lateral location closest to the separator in $\xcoord$-direction and near the edge of the NW or cathode particle in radial direction. The cathode SoC at EoC $\soc^\mathrm{c}_\mathrm{EoC}$ (squares) is solely included to prove that the Li supply from the cathode is never the limiting factor in this study ($\soc^\mathrm{c}_\mathrm{EoC} > 0.05$ for all cases). The anode SoC at EoC $\soc^\mathrm{a}_\mathrm{EoC}$ (circles) exhibits a plateau for $\porosity^\mathrm{a}_0 \geqq 0.65$. This reflects the standard charging behavior of the cell where the anode is fully charged. However, for $\porosity^\mathrm{a}_0 < 0.65$ the anode is not fully lithiated upon EoC and an increasing amount of Li remains stored in the cathode with decreasing initial anode porosity. From the depicted anode porosity at EoC $\porosity_\mathrm{EoC}^\mathrm{a}$ (crosses) we can see why. Decreasing the initial anode porosity naturally reduces the anode pore space left at the EoC. Thus, coming from high initial anode porosity $\porosity^\mathrm{a}_0$, the anode pore space at EoC $\porosity^\mathrm{a}_\mathrm{EoC}$ decreases until the geometry-related threshold of $\porosity^\mathrm{a}_0 = 0.22$ is reached at $\porosity^\mathrm{a}_0 = 0.65$. The depletion of anode pore space thus limits the charging capacity of the anode and the full cell for $\porosity^\mathrm{a}_0 < 0.65$ (\emph{cf.} \cref{fig:Cap_vs_eps_ano}).

Altogether, we conclude that, in the course of charging the cell, it is important for nanostructured Si anodes to provide enough pore space for the active material to expand into. By choosing the ideal initial anode porosity of 0.65, the cell capacity of the baseline cell can be improved by 60\% from \qty{1.0}{\milli \ampere \hour \per \cm \squared} (at an anode porosity of 0.78) to \qty{1.6}{\milli \ampere \hour \per \cm \squared}. This equals an active Si mass loading of \qty{0.4}{\mg \per \cm \squared} while maintaining the \qty{150}{\nm} initial radius of the NW (see \cref*{eq:SI_ActiveLoad} in the ESI).

\section{Conclusion}
\label{sec:Conclusion}

In this work, we describe and apply a physics-based 1d+1d modeling framework for Li-ion secondary batteries that incorporates a nanostructured Si anode and IL electrolyte. Our framework comprises important material-specific aspects of the transport. For the highly concentrated electrolyte, we include convection in the transport equations. On the anode side, the large volumetric changes of Si are taken into account through chemo-mechanical coupling.

We demonstrate the functionality of our model for a fully parameterized baseline cell with a Si nanowire (NW) anode and standard NMC111 cathode. Our physics-based simulations provide access to otherwise inaccessible quantities like the state of charge inside the active materials and the stresses that occur in the nanostructured Si.

We then perform parameter studies to investigate the influence of certain transport and geometric parameters of the Si NW anode on the cell performance. We find that for the low solid diffusion coefficient of Li in Si that is reported in literature nanostructured Si anodes are advantageous. Slow Li diffusion limits the capacity of the anode that can be exploited. Thicker Si structures do not get fully lithiated without risking Li plating. This leads to a reduced maximal achievable cell capacity for a constant active Si mass loading. Secondly, for nanostructured Si anodes sufficient pore space between the Si structures is important. Low porosity hinders the Si NWs from expanding and thus limits the maximum cell capacity.


\nomenclature[A,01]{a}{Superscript referring to anode}
\nomenclature[A,02]{c}{Superscript referring to cathode}
\nomenclature[A,03]{e}{Superscript referring to electrolyte}
\nomenclature[A,04]{H}{Superscript referring to Helmholtz}
\nomenclature[A,05]{max}{Superscript referring to maximum}
\nomenclature[A,06]{s}{Superscript referring to solid electrodes}
\nomenclature[A,09]{an}{Subscript referring to anion}
\nomenclature[A,10]{cat}{Subscript referring to cation}
\nomenclature[A,11]{chem}{Subscript referring to chemical}
\nomenclature[A,12]{elas}{Subscript referring to elastic}
\nomenclature[A,13]{EoC}{Subscript referring to End of Charge}
\nomenclature[A,14]{ext}{Subscript referring to external}
\nomenclature[A,16]{IL}{Subscript referring to ionic liquid (cation + anion)}
\nomenclature[A,17]{int}{Subscript referring to interaction}
\nomenclature[A,18]{Li}{Subscript referring to lithium}
\nomenclature[A,19]{LiSalt}{Subscript referring to Li-salt (Li + anion)}
\nomenclature[A,20]{r}{Subscript referring to radial component}
\nomenclature[A,22]{rev}{Subscript referring to reversible}
\nomenclature[A,23]{Si}{Subscript referring to silicon}
\nomenclature[A,24]{spec}{Subscript referring to specific}
\nomenclature[A,25]{tot}{Subscript referring to total}
\nomenclature[A,26]{$\alpha, \beta, \ldots$}{Greek subscripts refer to ion constituents}
\nomenclature[A,27]{$\upphi$}{Subscript referring to tangential/angular component}
\nomenclature[A,28]{0}{Subscript referring to initial/Lagrange frame}
\nomenclature[A,29]{ }{}

\nomenclature[B,01]{$\Load$}{Active mass loading \nomunit{\kilogram \per \meter \squared}}
\nomenclature[B,02]{$\Area$}{Area \nomunit{\meter \squared}}
\nomenclature[B,03]{$b$}{Body force density \nomunit{\meter \per \second \squared}}
\nomenclature[B,04]{$\conc$}{Species concentration / Molarity \nomunit{\mol \per \meter \cubed}}
\nomenclature[B,05]{$C$}{Capacity \nomunit{\ampere \second \per \meter \squared}}
\nomenclature[B,06]{$\CGreen$}{Reversible right Cauchy-Green tensor}
\nomenclature[B,07]{$\Elas$}{Elasticity tensor \nomunit{\kilogram \per \meter \per \second \squared}}
\nomenclature[B,08]{$\diffcoeff$}{Diffusion coefficient (with respect to thermodynamic driving force) \nomunit{\meter \squared \per \second}}
\nomenclature[B,09]{$\dieldis$}{Dielectric displacement field \nomunit{\ampere \second \per \meter \squared}}
\nomenclature[B,10]{$\Emodul$}{Young's modulus \nomunit{\kilogram \per \meter \per \second \squared}}
\nomenclature[B,11]{$\GreenLagr_\mathrm{elas}$}{Elastic Green-Lagrange strain tensor}
\nomenclature[B,13]{$\elfield$}{Electric field \nomunit{\kilogram \meter \per \ampere \per \second \cubed}}
\nomenclature[B,14]{$\Faraday$}{Faraday constant \nomunit{\ampere \second \per \mol}}
\nomenclature[B,15]{$\freeenergy$}{Helmholtz free energy \nomunit{\kilogram \meter \squared \per \second \squared}}
\nomenclature[B,16]{$\DeformGrad$}{Deformation gradient tensor}
\nomenclature[B,17]{$\elcurr$}{Electric current \nomunit{\ampere}}
\nomenclature[B,18]{$\mathbf{I}$}{Identity matrix}
\nomenclature[B,19]{$\ExCurrDens_0$}{Exchange current density \nomunit{\ampere \per \meter \squared}}
\nomenclature[B,20]{$\detF$}{Determinant of deformation gradient tensor}
\nomenclature[B,21]{$\Jflux$}{Electric current density \nomunit{\ampere \per \meter \squared}}
\nomenclature[B,22]{$\RateConst$}{Rate constant \nomunit{\mol \per \second \per \meter \squared}}
\nomenclature[B,23]{$\bulkmod$}{Bulk modulus \nomunit{\kilogram \per \meter \per \second \squared}}
\nomenclature[B,25]{$\Length$}{Length \nomunit{\meter}}
\nomenclature[B,26]{$\Onsager$}{Onsager matrix \nomunit{\second \mol \squared \per \kilogram \per \meter \cubed}}
\nomenclature[B,27]{$\mass$}{Molar mass \nomunit{\kilogram \per \mol}}
\nomenclature[B,28]{$\Mobility$}{Isotropic Li mobility \nomunit{\mol \squared \second \per \kilogram \per \meter \cubed}}
\nomenclature[B,29]{$\Number$}{Number of species}
\nomenclature[B,31]{$\Nflux$}{Species flux density \nomunit{\mol \per \meter \squared \per \second}}
\nomenclature[B,32]{$\momentum$}{Momentum density \nomunit{\meter \per \second}}
\nomenclature[B,33]{$\FirstPiola$}{First Piola-Kirchhoff stress tensor \nomunit{\kilogram \per \meter \per \second \squared}}
\nomenclature[B,34]{$\charge$}{Electric charge density \nomunit{\ampere \second \per \meter \cubed}}
\nomenclature[B,35]{$\rcoord$}{Radial coordinate (radius) \nomunit{\meter}}
\nomenclature[B,37]{$\Radius$}{Radius \nomunit{\meter}}
\nomenclature[B,38]{$\Rconst$}{Universal gas constant \nomunit{\kilogram \meter \squared \per \second \squared \per \kelvin \per \mol}}
\nomenclature[B,39]{$\entropyrate$}{Entropy production rate \nomunit{\kilogram \per \meter \per \second \cubed}}
\nomenclature[B,40]{$\source$}{Source term \nomunit{\mol \per \second \per \meter \cubed}}
\nomenclature[B,41]{$\entropy$}{Entropy density \nomunit{\meter \squared \per \second \squared \per \kelvin}}
\nomenclature[B,43]{$\Time$}{Time \nomunit{\second}}
\nomenclature[B,44]{$\transf_\alpha$}{Transference number}
\nomenclature[B,45]{$\Temp$}{Temperature \nomunit{\kelvin}}
\nomenclature[B,46]{$\SecondPiola$}{Second Piola-Kirchhoff stress tensor \nomunit{\kilogram \per \meter \per \second \squared}}
\nomenclature[B,47]{$\IntEnerg$}{Internal energy density \nomunit{\meter \squared \per \second \squared \per \kelvin}}
\nomenclature[B,49]{$\voltage$}{Potential \nomunit{\kilogram \meter \squared \per \ampere \per \second \cubed}}
\nomenclature[B,50]{$\Volume$}{Volume \nomunit{\meter \cubed}}
\nomenclature[B,51]{$\vel$}{(Convection) velocity (center of mass) \nomunit{\meter \per \second}}
\nomenclature[B,52]{$\xcoord$}{x coordinate (length) \nomunit{\meter}}
\nomenclature[B,53]{$\valence$}{Valence / Charge number}
\nomenclature[B,54]{ }{}

\nomenclature[C,01]{$\Brugge$}{Bruggeman coefficient}
\nomenclature[C,02]{$\porosity$}{Porosity / Volume fraction of pore space}
\nomenclature[C,03]{$\RelPermitt$}{Relative permittivity}
\nomenclature[C,04]{$\VacPermitt$}{Vacuum permittivity \nomunit{\ampere \second \per \volt \per \meter}}
\nomenclature[C,05]{$\OverPot$}{Overpotential \nomunit{\kilogram \meter \squared \per \ampere \per \second \cubed}}
\nomenclature[C,07]{$\soc$}{State of charge (SoC)}
\nomenclature[C,08]{$\conduc$}{Conductivity \nomunit{\ampere \squared \second \cubed \per \kilogram \per \meter \cubed}}
\nomenclature[C,09]{$\Deform$}{Deformation}
\nomenclature[C,10]{$\chempot$}{Chemical potential \nomunit{\kilogram \meter \squared \per \second \squared \per \mol}}
\nomenclature[C,11]{$\Poisson$}{Poisson ratio}
\nomenclature[C,13]{$\pmv$}{Partial molar volume \nomunit{\meter \cubed \per \mol}}
\nomenclature[C,14]{$\dens$}{Mass density \nomunit{\kilogram \per \meter \cubed}}
\nomenclature[C,15]{$\Cauchy$}{Cauchy stress tensor \nomunit{\kilogram \per \meter \per \second \squared}}
\nomenclature[C,16]{$\elpot$}{Electric potential \nomunit{\kilogram \meter \squared \per \ampere \per \second \cubed}}
\nomenclature[C,17]{$\altelpot$}{Chemo-electric potential \nomunit{\kilogram \meter \squared \per \ampere \per \second \cubed}}
\nomenclature[C,19]{$\varphi$}{Electrochemical potential \nomunit{\kilogram \meter \squared \per \second \squared \per \mol}}
\nomenclature[C,20]{$\freeenergydens$}{Helmholtz free energy density \nomunit{\meter \squared \per \second \squared}}

\printnomenclature

\section*{Acknowledgment}
This work was supported by the European Union's Horizon 2020 research and innovation program via the "Si-DRIVE" project (grant agreement No 814464).
This work was supported by the European Union's Horizon Europe via the research initiative Battery 2030+ via the "OPINCHARGE" project (grant agreement No 101104032).
The authors thank Lars von Kolzenberg, Lukas Köbbing and Linda Bolay for fruitful discussions.

\section*{Conflict of Interest}
The authors declare no conflict of interest.

\bibliographystyle{apsrev4-2}
\bibliography{bibliography}

\begin{thebibliography}{69}%
\makeatletter
\providecommand \@ifxundefined [1]{%
 \@ifx{#1\undefined}
}%
\providecommand \@ifnum [1]{%
 \ifnum #1\expandafter \@firstoftwo
 \else \expandafter \@secondoftwo
 \fi
}%
\providecommand \@ifx [1]{%
 \ifx #1\expandafter \@firstoftwo
 \else \expandafter \@secondoftwo
 \fi
}%
\providecommand \natexlab [1]{#1}%
\providecommand \enquote  [1]{``#1''}%
\providecommand \bibnamefont  [1]{#1}%
\providecommand \bibfnamefont [1]{#1}%
\providecommand \citenamefont [1]{#1}%
\providecommand \href@noop [0]{\@secondoftwo}%
\providecommand \href [0]{\begingroup \@sanitize@url \@href}%
\providecommand \@href[1]{\@@startlink{#1}\@@href}%
\providecommand \@@href[1]{\endgroup#1\@@endlink}%
\providecommand \@sanitize@url [0]{\catcode `\\12\catcode `\$12\catcode
  `\&12\catcode `\#12\catcode `\^12\catcode `\_12\catcode `\%12\relax}%
\providecommand \@@startlink[1]{}%
\providecommand \@@endlink[0]{}%
\providecommand \url  [0]{\begingroup\@sanitize@url \@url }%
\providecommand \@url [1]{\endgroup\@href {#1}{\urlprefix }}%
\providecommand \urlprefix  [0]{URL }%
\providecommand \Eprint [0]{\href }%
\providecommand \doibase [0]{https://doi.org/}%
\providecommand \selectlanguage [0]{\@gobble}%
\providecommand \bibinfo  [0]{\@secondoftwo}%
\providecommand \bibfield  [0]{\@secondoftwo}%
\providecommand \translation [1]{[#1]}%
\providecommand \BibitemOpen [0]{}%
\providecommand \bibitemStop [0]{}%
\providecommand \bibitemNoStop [0]{.\EOS\space}%
\providecommand \EOS [0]{\spacefactor3000\relax}%
\providecommand \BibitemShut  [1]{\csname bibitem#1\endcsname}%
\let\auto@bib@innerbib\@empty
\bibitem [{\citenamefont {Obrovac}\ and\ \citenamefont
  {Christensen}(2004)}]{Obrovac2004}%
  \BibitemOpen
  \bibfield  {author} {\bibinfo {author} {\bibfnamefont {M.~N.}\ \bibnamefont
  {Obrovac}}\ and\ \bibinfo {author} {\bibfnamefont {L.}~\bibnamefont
  {Christensen}},\ }\href {https://doi.org/10.1149/1.1652421} {\bibfield
  {journal} {\bibinfo  {journal} {Electrochemical and Solid-State Letters}\
  }\textbf {\bibinfo {volume} {7}},\ \bibinfo {pages} {A93} (\bibinfo {year}
  {2004})}\BibitemShut {NoStop}%
\bibitem [{\citenamefont {Obrovac}\ and\ \citenamefont
  {Krause}(2007)}]{Obrovac2007a}%
  \BibitemOpen
  \bibfield  {author} {\bibinfo {author} {\bibfnamefont {M.~N.}\ \bibnamefont
  {Obrovac}}\ and\ \bibinfo {author} {\bibfnamefont {L.~J.}\ \bibnamefont
  {Krause}},\ }\href {https://doi.org/10.1149/1.2402112} {\bibfield  {journal}
  {\bibinfo  {journal} {Journal of The Electrochemical Society}\ }\textbf
  {\bibinfo {volume} {154}},\ \bibinfo {pages} {A103} (\bibinfo {year}
  {2007})}\BibitemShut {NoStop}%
\bibitem [{\citenamefont {Winter}\ \emph {et~al.}(1998)\citenamefont {Winter},
  \citenamefont {Besenhard}, \citenamefont {Spahr},\ and\ \citenamefont
  {Novák}}]{Winter1998}%
  \BibitemOpen
  \bibfield  {author} {\bibinfo {author} {\bibfnamefont {M.}~\bibnamefont
  {Winter}}, \bibinfo {author} {\bibfnamefont {J.~O.}\ \bibnamefont
  {Besenhard}}, \bibinfo {author} {\bibfnamefont {M.~E.}\ \bibnamefont
  {Spahr}},\ and\ \bibinfo {author} {\bibfnamefont {P.}~\bibnamefont
  {Novák}},\ }\href
  {https://doi.org/10.1002/(SICI)1521-4095(199807)10:10<725::AID-ADMA725>3.0.CO;2-Z}
  {\bibfield  {journal} {\bibinfo  {journal} {Advanced Materials}\ }\textbf
  {\bibinfo {volume} {10}},\ \bibinfo {pages} {725} (\bibinfo {year}
  {1998})}\BibitemShut {NoStop}%
\bibitem [{\citenamefont {Soni}\ \emph {et~al.}(2012)\citenamefont {Soni},
  \citenamefont {Sheldon}, \citenamefont {Xiao}, \citenamefont {Bower},\ and\
  \citenamefont {Verbrugge}}]{Soni2012}%
  \BibitemOpen
  \bibfield  {author} {\bibinfo {author} {\bibfnamefont {S.~K.}\ \bibnamefont
  {Soni}}, \bibinfo {author} {\bibfnamefont {B.~W.}\ \bibnamefont {Sheldon}},
  \bibinfo {author} {\bibfnamefont {X.}~\bibnamefont {Xiao}}, \bibinfo {author}
  {\bibfnamefont {A.~F.}\ \bibnamefont {Bower}},\ and\ \bibinfo {author}
  {\bibfnamefont {M.~W.}\ \bibnamefont {Verbrugge}},\ }\href
  {https://doi.org/10.1149/2.009209jes} {\bibfield  {journal} {\bibinfo
  {journal} {Journal of The Electrochemical Society}\ }\textbf {\bibinfo
  {volume} {159}},\ \bibinfo {pages} {A1520} (\bibinfo {year}
  {2012})}\BibitemShut {NoStop}%
\bibitem [{\citenamefont {von Kolzenberg}\ \emph {et~al.}(2022)\citenamefont
  {von Kolzenberg}, \citenamefont {Latz},\ and\ \citenamefont
  {Horstmann}}]{Kolzenberg2022a}%
  \BibitemOpen
  \bibfield  {author} {\bibinfo {author} {\bibfnamefont {L.}~\bibnamefont {von
  Kolzenberg}}, \bibinfo {author} {\bibfnamefont {A.}~\bibnamefont {Latz}},\
  and\ \bibinfo {author} {\bibfnamefont {B.}~\bibnamefont {Horstmann}},\
  }\bibfield  {journal} {\bibinfo  {journal} {Batteries \& Supercaps}\ }\textbf
  {\bibinfo {volume} {5}},\ \href {https://doi.org/10.1002/batt.202100216}
  {10.1002/batt.202100216} (\bibinfo {year} {2022})\BibitemShut {NoStop}%
\bibitem [{\citenamefont {Zilberman}\ \emph {et~al.}(2019)\citenamefont
  {Zilberman}, \citenamefont {Sturm},\ and\ \citenamefont
  {Jossen}}]{Zilberman2019}%
  \BibitemOpen
  \bibfield  {author} {\bibinfo {author} {\bibfnamefont {I.}~\bibnamefont
  {Zilberman}}, \bibinfo {author} {\bibfnamefont {J.}~\bibnamefont {Sturm}},\
  and\ \bibinfo {author} {\bibfnamefont {A.}~\bibnamefont {Jossen}},\ }\href
  {https://doi.org/10.1016/j.jpowsour.2019.03.109} {\bibfield  {journal}
  {\bibinfo  {journal} {Journal of Power Sources}\ }\textbf {\bibinfo {volume}
  {425}},\ \bibinfo {pages} {217} (\bibinfo {year} {2019})}\BibitemShut
  {NoStop}%
\bibitem [{\citenamefont {Li}\ \emph {et~al.}(2019)\citenamefont {Li},
  \citenamefont {Liu}, \citenamefont {Kong}, \citenamefont {Cheng},\ and\
  \citenamefont {Zhao}}]{Li2019}%
  \BibitemOpen
  \bibfield  {author} {\bibinfo {author} {\bibfnamefont {H.}~\bibnamefont
  {Li}}, \bibinfo {author} {\bibfnamefont {C.}~\bibnamefont {Liu}}, \bibinfo
  {author} {\bibfnamefont {X.}~\bibnamefont {Kong}}, \bibinfo {author}
  {\bibfnamefont {J.}~\bibnamefont {Cheng}},\ and\ \bibinfo {author}
  {\bibfnamefont {J.}~\bibnamefont {Zhao}},\ }\href
  {https://doi.org/10.1016/j.jpowsour.2019.226971} {\bibfield  {journal}
  {\bibinfo  {journal} {Journal of Power Sources}\ }\textbf {\bibinfo {volume}
  {438}},\ \bibinfo {pages} {226971} (\bibinfo {year} {2019})}\BibitemShut
  {NoStop}%
\bibitem [{\citenamefont {Anseán}\ \emph {et~al.}(2020)\citenamefont
  {Anseán}, \citenamefont {Baure}, \citenamefont {González}, \citenamefont
  {Cameán}, \citenamefont {García},\ and\ \citenamefont
  {Dubarry}}]{Ansean2020}%
  \BibitemOpen
  \bibfield  {author} {\bibinfo {author} {\bibfnamefont {D.}~\bibnamefont
  {Anseán}}, \bibinfo {author} {\bibfnamefont {G.}~\bibnamefont {Baure}},
  \bibinfo {author} {\bibfnamefont {M.}~\bibnamefont {González}}, \bibinfo
  {author} {\bibfnamefont {I.}~\bibnamefont {Cameán}}, \bibinfo {author}
  {\bibfnamefont {A.}~\bibnamefont {García}},\ and\ \bibinfo {author}
  {\bibfnamefont {M.}~\bibnamefont {Dubarry}},\ }\href
  {https://doi.org/10.1016/j.jpowsour.2020.227882} {\bibfield  {journal}
  {\bibinfo  {journal} {Journal of Power Sources}\ }\textbf {\bibinfo {volume}
  {459}},\ \bibinfo {pages} {227882} (\bibinfo {year} {2020})}\BibitemShut
  {NoStop}%
\bibitem [{\citenamefont {Chen}\ \emph {et~al.}(2022)\citenamefont {Chen},
  \citenamefont {Danilov}, \citenamefont {Zhang}, \citenamefont {Jiang},
  \citenamefont {Zhou}, \citenamefont {Eichel},\ and\ \citenamefont
  {Notten}}]{Chen2022}%
  \BibitemOpen
  \bibfield  {author} {\bibinfo {author} {\bibfnamefont {Z.}~\bibnamefont
  {Chen}}, \bibinfo {author} {\bibfnamefont {D.~L.}\ \bibnamefont {Danilov}},
  \bibinfo {author} {\bibfnamefont {Q.}~\bibnamefont {Zhang}}, \bibinfo
  {author} {\bibfnamefont {M.}~\bibnamefont {Jiang}}, \bibinfo {author}
  {\bibfnamefont {J.}~\bibnamefont {Zhou}}, \bibinfo {author} {\bibfnamefont
  {R.-A.}\ \bibnamefont {Eichel}},\ and\ \bibinfo {author} {\bibfnamefont
  {P.~H.}\ \bibnamefont {Notten}},\ }\href
  {https://doi.org/10.1016/j.electacta.2022.141077} {\bibfield  {journal}
  {\bibinfo  {journal} {Electrochimica Acta}\ }\textbf {\bibinfo {volume}
  {430}},\ \bibinfo {pages} {141077} (\bibinfo {year} {2022})}\BibitemShut
  {NoStop}%
\bibitem [{\citenamefont {Szczech}\ and\ \citenamefont
  {Jin}(2011)}]{Szczech2011}%
  \BibitemOpen
  \bibfield  {author} {\bibinfo {author} {\bibfnamefont {J.~R.}\ \bibnamefont
  {Szczech}}\ and\ \bibinfo {author} {\bibfnamefont {S.}~\bibnamefont {Jin}},\
  }\href {https://doi.org/10.1039/C0EE00281J} {\bibfield  {journal} {\bibinfo
  {journal} {Energy \& Environmental Science}\ }\textbf {\bibinfo {volume}
  {4}},\ \bibinfo {pages} {56} (\bibinfo {year} {2011})}\BibitemShut {NoStop}%
\bibitem [{\citenamefont {Wu}\ and\ \citenamefont {Cui}(2012)}]{Wu2012}%
  \BibitemOpen
  \bibfield  {author} {\bibinfo {author} {\bibfnamefont {H.}~\bibnamefont
  {Wu}}\ and\ \bibinfo {author} {\bibfnamefont {Y.}~\bibnamefont {Cui}},\
  }\href {https://doi.org/10.1016/j.nantod.2012.08.004} {\bibfield  {journal}
  {\bibinfo  {journal} {Nano Today}\ }\textbf {\bibinfo {volume} {7}},\
  \bibinfo {pages} {414} (\bibinfo {year} {2012})}\BibitemShut {NoStop}%
\bibitem [{\citenamefont {Rahman}\ \emph {et~al.}(2016)\citenamefont {Rahman},
  \citenamefont {Song}, \citenamefont {Bhatt}, \citenamefont {Wong},\ and\
  \citenamefont {Wen}}]{Rahman2016}%
  \BibitemOpen
  \bibfield  {author} {\bibinfo {author} {\bibfnamefont {M.~A.}\ \bibnamefont
  {Rahman}}, \bibinfo {author} {\bibfnamefont {G.}~\bibnamefont {Song}},
  \bibinfo {author} {\bibfnamefont {A.~I.}\ \bibnamefont {Bhatt}}, \bibinfo
  {author} {\bibfnamefont {Y.~C.}\ \bibnamefont {Wong}},\ and\ \bibinfo
  {author} {\bibfnamefont {C.}~\bibnamefont {Wen}},\ }\href
  {https://doi.org/10.1002/adfm.201502959} {\bibfield  {journal} {\bibinfo
  {journal} {Advanced Functional Materials}\ }\textbf {\bibinfo {volume}
  {26}},\ \bibinfo {pages} {647} (\bibinfo {year} {2016})}\BibitemShut
  {NoStop}%
\bibitem [{\citenamefont {Cheng}\ and\ \citenamefont
  {Verbrugge}(2010)}]{Cheng2010}%
  \BibitemOpen
  \bibfield  {author} {\bibinfo {author} {\bibfnamefont {Y.-T.}\ \bibnamefont
  {Cheng}}\ and\ \bibinfo {author} {\bibfnamefont {M.~W.}\ \bibnamefont
  {Verbrugge}},\ }\href {https://doi.org/10.1149/1.3298892} {\bibfield
  {journal} {\bibinfo  {journal} {Journal of The Electrochemical Society}\
  }\textbf {\bibinfo {volume} {157}},\ \bibinfo {pages} {A508} (\bibinfo {year}
  {2010})}\BibitemShut {NoStop}%
\bibitem [{\citenamefont {Deshpande}\ \emph {et~al.}(2010)\citenamefont
  {Deshpande}, \citenamefont {Cheng},\ and\ \citenamefont
  {Verbrugge}}]{Deshpande2010}%
  \BibitemOpen
  \bibfield  {author} {\bibinfo {author} {\bibfnamefont {R.}~\bibnamefont
  {Deshpande}}, \bibinfo {author} {\bibfnamefont {Y.-T.}\ \bibnamefont
  {Cheng}},\ and\ \bibinfo {author} {\bibfnamefont {M.~W.}\ \bibnamefont
  {Verbrugge}},\ }\href {https://doi.org/10.1016/j.jpowsour.2010.02.021}
  {\bibfield  {journal} {\bibinfo  {journal} {Journal of Power Sources}\
  }\textbf {\bibinfo {volume} {195}},\ \bibinfo {pages} {5081} (\bibinfo {year}
  {2010})}\BibitemShut {NoStop}%
\bibitem [{\citenamefont {Ryu}\ \emph {et~al.}(2011)\citenamefont {Ryu},
  \citenamefont {Choi}, \citenamefont {Cui},\ and\ \citenamefont
  {Nix}}]{Ryu2011}%
  \BibitemOpen
  \bibfield  {author} {\bibinfo {author} {\bibfnamefont {I.}~\bibnamefont
  {Ryu}}, \bibinfo {author} {\bibfnamefont {J.~W.}\ \bibnamefont {Choi}},
  \bibinfo {author} {\bibfnamefont {Y.}~\bibnamefont {Cui}},\ and\ \bibinfo
  {author} {\bibfnamefont {W.~D.}\ \bibnamefont {Nix}},\ }\href
  {https://doi.org/10.1016/j.jmps.2011.06.003} {\bibfield  {journal} {\bibinfo
  {journal} {Journal of the Mechanics and Physics of Solids}\ }\textbf
  {\bibinfo {volume} {59}},\ \bibinfo {pages} {1717} (\bibinfo {year}
  {2011})}\BibitemShut {NoStop}%
\bibitem [{\citenamefont {Liu}\ \emph {et~al.}(2012)\citenamefont {Liu},
  \citenamefont {Zhong}, \citenamefont {Huang}, \citenamefont {Mao},
  \citenamefont {Zhu},\ and\ \citenamefont {Huang}}]{Liu2012a}%
  \BibitemOpen
  \bibfield  {author} {\bibinfo {author} {\bibfnamefont {X.~H.}\ \bibnamefont
  {Liu}}, \bibinfo {author} {\bibfnamefont {L.}~\bibnamefont {Zhong}}, \bibinfo
  {author} {\bibfnamefont {S.}~\bibnamefont {Huang}}, \bibinfo {author}
  {\bibfnamefont {S.~X.}\ \bibnamefont {Mao}}, \bibinfo {author} {\bibfnamefont
  {T.}~\bibnamefont {Zhu}},\ and\ \bibinfo {author} {\bibfnamefont {J.~Y.}\
  \bibnamefont {Huang}},\ }\href {https://doi.org/10.1021/nn204476h} {\bibfield
   {journal} {\bibinfo  {journal} {ACS Nano}\ }\textbf {\bibinfo {volume}
  {6}},\ \bibinfo {pages} {1522} (\bibinfo {year} {2012})}\BibitemShut
  {NoStop}%
\bibitem [{\citenamefont {Kennedy}\ \emph {et~al.}(2016)\citenamefont
  {Kennedy}, \citenamefont {Brandon},\ and\ \citenamefont
  {Ryan}}]{Kennedy2016}%
  \BibitemOpen
  \bibfield  {author} {\bibinfo {author} {\bibfnamefont {T.}~\bibnamefont
  {Kennedy}}, \bibinfo {author} {\bibfnamefont {M.}~\bibnamefont {Brandon}},\
  and\ \bibinfo {author} {\bibfnamefont {K.~M.}\ \bibnamefont {Ryan}},\ }\href
  {https://doi.org/10.1002/adma.201503978} {\bibfield  {journal} {\bibinfo
  {journal} {Advanced Materials}\ }\textbf {\bibinfo {volume} {28}},\ \bibinfo
  {pages} {5696} (\bibinfo {year} {2016})}\BibitemShut {NoStop}%
\bibitem [{\citenamefont {Zhou}\ \emph {et~al.}(2019)\citenamefont {Zhou},
  \citenamefont {Xu}, \citenamefont {Hu}, \citenamefont {Mai},\ and\
  \citenamefont {Cui}}]{Zhou2019}%
  \BibitemOpen
  \bibfield  {author} {\bibinfo {author} {\bibfnamefont {G.}~\bibnamefont
  {Zhou}}, \bibinfo {author} {\bibfnamefont {L.}~\bibnamefont {Xu}}, \bibinfo
  {author} {\bibfnamefont {G.}~\bibnamefont {Hu}}, \bibinfo {author}
  {\bibfnamefont {L.}~\bibnamefont {Mai}},\ and\ \bibinfo {author}
  {\bibfnamefont {Y.}~\bibnamefont {Cui}},\ }\href
  {https://doi.org/10.1021/acs.chemrev.9b00326} {\bibfield  {journal} {\bibinfo
   {journal} {Chemical Reviews}\ }\textbf {\bibinfo {volume} {119}},\ \bibinfo
  {pages} {11042} (\bibinfo {year} {2019})}\BibitemShut {NoStop}%
\bibitem [{\citenamefont {Chan}\ \emph {et~al.}(2008)\citenamefont {Chan},
  \citenamefont {Peng}, \citenamefont {Liu}, \citenamefont {McIlwrath},
  \citenamefont {Zhang}, \citenamefont {Huggins},\ and\ \citenamefont
  {Cui}}]{Chan2008}%
  \BibitemOpen
  \bibfield  {author} {\bibinfo {author} {\bibfnamefont {C.~K.}\ \bibnamefont
  {Chan}}, \bibinfo {author} {\bibfnamefont {H.}~\bibnamefont {Peng}}, \bibinfo
  {author} {\bibfnamefont {G.}~\bibnamefont {Liu}}, \bibinfo {author}
  {\bibfnamefont {K.}~\bibnamefont {McIlwrath}}, \bibinfo {author}
  {\bibfnamefont {X.~F.}\ \bibnamefont {Zhang}}, \bibinfo {author}
  {\bibfnamefont {R.~A.}\ \bibnamefont {Huggins}},\ and\ \bibinfo {author}
  {\bibfnamefont {Y.}~\bibnamefont {Cui}},\ }\href
  {https://doi.org/10.1038/nnano.2007.411} {\bibfield  {journal} {\bibinfo
  {journal} {Nature Nanotechnology}\ }\textbf {\bibinfo {volume} {3}},\
  \bibinfo {pages} {31} (\bibinfo {year} {2008})}\BibitemShut {NoStop}%
\bibitem [{\citenamefont {Cui}\ \emph {et~al.}(2009)\citenamefont {Cui},
  \citenamefont {Ruffo}, \citenamefont {Chan}, \citenamefont {Peng},\ and\
  \citenamefont {Cui}}]{Cui2009}%
  \BibitemOpen
  \bibfield  {author} {\bibinfo {author} {\bibfnamefont {L.-F.}\ \bibnamefont
  {Cui}}, \bibinfo {author} {\bibfnamefont {R.}~\bibnamefont {Ruffo}}, \bibinfo
  {author} {\bibfnamefont {C.~K.}\ \bibnamefont {Chan}}, \bibinfo {author}
  {\bibfnamefont {H.}~\bibnamefont {Peng}},\ and\ \bibinfo {author}
  {\bibfnamefont {Y.}~\bibnamefont {Cui}},\ }\href
  {https://doi.org/10.1021/nl8036323} {\bibfield  {journal} {\bibinfo
  {journal} {Nano Letters}\ }\textbf {\bibinfo {volume} {9}},\ \bibinfo {pages}
  {491} (\bibinfo {year} {2009})}\BibitemShut {NoStop}%
\bibitem [{\citenamefont {Zamfir}\ \emph {et~al.}(2013)\citenamefont {Zamfir},
  \citenamefont {Nguyen}, \citenamefont {Moyen}, \citenamefont {Lee},\ and\
  \citenamefont {Pribat}}]{Zamfir2013}%
  \BibitemOpen
  \bibfield  {author} {\bibinfo {author} {\bibfnamefont {M.~R.}\ \bibnamefont
  {Zamfir}}, \bibinfo {author} {\bibfnamefont {H.~T.}\ \bibnamefont {Nguyen}},
  \bibinfo {author} {\bibfnamefont {E.}~\bibnamefont {Moyen}}, \bibinfo
  {author} {\bibfnamefont {Y.~H.}\ \bibnamefont {Lee}},\ and\ \bibinfo {author}
  {\bibfnamefont {D.}~\bibnamefont {Pribat}},\ }\href
  {https://doi.org/10.1039/c3ta11714f} {\bibfield  {journal} {\bibinfo
  {journal} {Journal of Materials Chemistry A}\ }\textbf {\bibinfo {volume}
  {1}},\ \bibinfo {pages} {9566} (\bibinfo {year} {2013})}\BibitemShut
  {NoStop}%
\bibitem [{\citenamefont {Yang}\ \emph {et~al.}(2020)\citenamefont {Yang},
  \citenamefont {Yuan}, \citenamefont {Kang}, \citenamefont {Ye}, \citenamefont
  {Pan}, \citenamefont {Zhang}, \citenamefont {Ke}, \citenamefont {Wang},
  \citenamefont {Qiu},\ and\ \citenamefont {Tang}}]{Yang2020}%
  \BibitemOpen
  \bibfield  {author} {\bibinfo {author} {\bibfnamefont {Y.}~\bibnamefont
  {Yang}}, \bibinfo {author} {\bibfnamefont {W.}~\bibnamefont {Yuan}}, \bibinfo
  {author} {\bibfnamefont {W.}~\bibnamefont {Kang}}, \bibinfo {author}
  {\bibfnamefont {Y.}~\bibnamefont {Ye}}, \bibinfo {author} {\bibfnamefont
  {Q.}~\bibnamefont {Pan}}, \bibinfo {author} {\bibfnamefont {X.}~\bibnamefont
  {Zhang}}, \bibinfo {author} {\bibfnamefont {Y.}~\bibnamefont {Ke}}, \bibinfo
  {author} {\bibfnamefont {C.}~\bibnamefont {Wang}}, \bibinfo {author}
  {\bibfnamefont {Z.}~\bibnamefont {Qiu}},\ and\ \bibinfo {author}
  {\bibfnamefont {Y.}~\bibnamefont {Tang}},\ }\href
  {https://doi.org/10.1039/C9SE01165J} {\bibfield  {journal} {\bibinfo
  {journal} {Sustainable Energy \& Fuels}\ }\textbf {\bibinfo {volume} {4}},\
  \bibinfo {pages} {1577} (\bibinfo {year} {2020})}\BibitemShut {NoStop}%
\bibitem [{\citenamefont {Chen}\ \emph {et~al.}(2014)\citenamefont {Chen},
  \citenamefont {Fan}, \citenamefont {Hong}, \citenamefont {Chen},
  \citenamefont {Ji}, \citenamefont {Zhang}, \citenamefont {Zhu},\ and\
  \citenamefont {Chen}}]{Chen2014}%
  \BibitemOpen
  \bibfield  {author} {\bibinfo {author} {\bibfnamefont {L.}~\bibnamefont
  {Chen}}, \bibinfo {author} {\bibfnamefont {F.}~\bibnamefont {Fan}}, \bibinfo
  {author} {\bibfnamefont {L.}~\bibnamefont {Hong}}, \bibinfo {author}
  {\bibfnamefont {J.}~\bibnamefont {Chen}}, \bibinfo {author} {\bibfnamefont
  {Y.~Z.}\ \bibnamefont {Ji}}, \bibinfo {author} {\bibfnamefont {S.~L.}\
  \bibnamefont {Zhang}}, \bibinfo {author} {\bibfnamefont {T.}~\bibnamefont
  {Zhu}},\ and\ \bibinfo {author} {\bibfnamefont {L.~Q.}\ \bibnamefont
  {Chen}},\ }\href {https://doi.org/10.1149/2.0171411jes} {\bibfield  {journal}
  {\bibinfo  {journal} {Journal of The Electrochemical Society}\ }\textbf
  {\bibinfo {volume} {161}},\ \bibinfo {pages} {F3164} (\bibinfo {year}
  {2014})}\BibitemShut {NoStop}%
\bibitem [{\citenamefont {Xie}\ \emph {et~al.}(2015)\citenamefont {Xie},
  \citenamefont {Qiu}, \citenamefont {Gao}, \citenamefont {Guan},\ and\
  \citenamefont {Yuan}}]{Xie2015}%
  \BibitemOpen
  \bibfield  {author} {\bibinfo {author} {\bibfnamefont {Y.}~\bibnamefont
  {Xie}}, \bibinfo {author} {\bibfnamefont {M.}~\bibnamefont {Qiu}}, \bibinfo
  {author} {\bibfnamefont {X.}~\bibnamefont {Gao}}, \bibinfo {author}
  {\bibfnamefont {D.}~\bibnamefont {Guan}},\ and\ \bibinfo {author}
  {\bibfnamefont {C.}~\bibnamefont {Yuan}},\ }\href
  {https://doi.org/10.1016/j.electacta.2015.11.022} {\bibfield  {journal}
  {\bibinfo  {journal} {Electrochimica Acta}\ }\textbf {\bibinfo {volume}
  {186}},\ \bibinfo {pages} {542} (\bibinfo {year} {2015})}\BibitemShut
  {NoStop}%
\bibitem [{\citenamefont {Zuo}\ and\ \citenamefont {Zhao}(2015)}]{Zuo2015}%
  \BibitemOpen
  \bibfield  {author} {\bibinfo {author} {\bibfnamefont {P.}~\bibnamefont
  {Zuo}}\ and\ \bibinfo {author} {\bibfnamefont {Y.-P.}\ \bibnamefont {Zhao}},\
  }\href {https://doi.org/10.1039/C4CP00563E} {\bibfield  {journal} {\bibinfo
  {journal} {Physical Chemistry Chemical Physics}\ }\textbf {\bibinfo {volume}
  {17}},\ \bibinfo {pages} {287} (\bibinfo {year} {2015})}\BibitemShut
  {NoStop}%
\bibitem [{\citenamefont {Gao}\ and\ \citenamefont {Hong}(2016)}]{Gao2016}%
  \BibitemOpen
  \bibfield  {author} {\bibinfo {author} {\bibfnamefont {F.}~\bibnamefont
  {Gao}}\ and\ \bibinfo {author} {\bibfnamefont {W.}~\bibnamefont {Hong}},\
  }\href {https://doi.org/10.1016/j.jmps.2016.04.020} {\bibfield  {journal}
  {\bibinfo  {journal} {Journal of the Mechanics and Physics of Solids}\
  }\textbf {\bibinfo {volume} {94}},\ \bibinfo {pages} {18} (\bibinfo {year}
  {2016})}\BibitemShut {NoStop}%
\bibitem [{\citenamefont {Castelli}\ \emph {et~al.}(2021)\citenamefont
  {Castelli}, \citenamefont {von Kolzenberg}, \citenamefont {Horstmann},
  \citenamefont {Latz},\ and\ \citenamefont {Dörfler}}]{Castelli2021}%
  \BibitemOpen
  \bibfield  {author} {\bibinfo {author} {\bibfnamefont {G.~F.}\ \bibnamefont
  {Castelli}}, \bibinfo {author} {\bibfnamefont {L.}~\bibnamefont {von
  Kolzenberg}}, \bibinfo {author} {\bibfnamefont {B.}~\bibnamefont
  {Horstmann}}, \bibinfo {author} {\bibfnamefont {A.}~\bibnamefont {Latz}},\
  and\ \bibinfo {author} {\bibfnamefont {W.}~\bibnamefont {Dörfler}},\ }\href
  {https://doi.org/10.1002/ente.202000835} {\bibfield  {journal} {\bibinfo
  {journal} {Energy Technology}\ }\textbf {\bibinfo {volume} {9}},\ \bibinfo
  {pages} {2000835} (\bibinfo {year} {2021})}\BibitemShut {NoStop}%
\bibitem [{\citenamefont {Sikha}\ \emph {et~al.}(2014)\citenamefont {Sikha},
  \citenamefont {De},\ and\ \citenamefont {Gordon}}]{Sikha2014}%
  \BibitemOpen
  \bibfield  {author} {\bibinfo {author} {\bibfnamefont {G.}~\bibnamefont
  {Sikha}}, \bibinfo {author} {\bibfnamefont {S.}~\bibnamefont {De}},\ and\
  \bibinfo {author} {\bibfnamefont {J.}~\bibnamefont {Gordon}},\ }\href
  {https://doi.org/10.1016/j.jpowsour.2014.03.022} {\bibfield  {journal}
  {\bibinfo  {journal} {Journal of Power Sources}\ }\textbf {\bibinfo {volume}
  {262}},\ \bibinfo {pages} {514} (\bibinfo {year} {2014})}\BibitemShut
  {NoStop}%
\bibitem [{\citenamefont {De}\ \emph {et~al.}(2014)\citenamefont {De},
  \citenamefont {Gordon},\ and\ \citenamefont {Sikha}}]{De2014}%
  \BibitemOpen
  \bibfield  {author} {\bibinfo {author} {\bibfnamefont {S.}~\bibnamefont
  {De}}, \bibinfo {author} {\bibfnamefont {J.}~\bibnamefont {Gordon}},\ and\
  \bibinfo {author} {\bibfnamefont {G.}~\bibnamefont {Sikha}},\ }\href
  {https://doi.org/10.1016/j.jpowsour.2014.03.130} {\bibfield  {journal}
  {\bibinfo  {journal} {Journal of Power Sources}\ }\textbf {\bibinfo {volume}
  {262}},\ \bibinfo {pages} {524} (\bibinfo {year} {2014})}\BibitemShut
  {NoStop}%
\bibitem [{\citenamefont {Yang}\ \emph {et~al.}(2014)\citenamefont {Yang},
  \citenamefont {Fan}, \citenamefont {Liang}, \citenamefont {Guo},
  \citenamefont {Zhu},\ and\ \citenamefont {Zhang}}]{Yang2014}%
  \BibitemOpen
  \bibfield  {author} {\bibinfo {author} {\bibfnamefont {H.}~\bibnamefont
  {Yang}}, \bibinfo {author} {\bibfnamefont {F.}~\bibnamefont {Fan}}, \bibinfo
  {author} {\bibfnamefont {W.}~\bibnamefont {Liang}}, \bibinfo {author}
  {\bibfnamefont {X.}~\bibnamefont {Guo}}, \bibinfo {author} {\bibfnamefont
  {T.}~\bibnamefont {Zhu}},\ and\ \bibinfo {author} {\bibfnamefont
  {S.}~\bibnamefont {Zhang}},\ }\href
  {https://doi.org/10.1016/j.jmps.2014.06.004} {\bibfield  {journal} {\bibinfo
  {journal} {Journal of the Mechanics and Physics of Solids}\ }\textbf
  {\bibinfo {volume} {70}},\ \bibinfo {pages} {349} (\bibinfo {year}
  {2014})}\BibitemShut {NoStop}%
\bibitem [{\citenamefont {Köbbing}\ \emph {et~al.}(2023)\citenamefont
  {Köbbing}, \citenamefont {Latz},\ and\ \citenamefont
  {Horstmann}}]{Koebbing2023}%
  \BibitemOpen
  \bibfield  {author} {\bibinfo {author} {\bibfnamefont {L.}~\bibnamefont
  {Köbbing}}, \bibinfo {author} {\bibfnamefont {A.}~\bibnamefont {Latz}},\
  and\ \bibinfo {author} {\bibfnamefont {B.}~\bibnamefont {Horstmann}},\
  }\bibfield  {journal} {\bibinfo  {journal} {Advanced Functional Materials}\
  }\href {https://doi.org/10.1002/adfm.202308818} {10.1002/adfm.202308818}
  (\bibinfo {year} {2023})\BibitemShut {NoStop}%
\bibitem [{\citenamefont {Verma}\ \emph {et~al.}(2019)\citenamefont {Verma},
  \citenamefont {Franco},\ and\ \citenamefont {Mukherjee}}]{Verma2019}%
  \BibitemOpen
  \bibfield  {author} {\bibinfo {author} {\bibfnamefont {A.}~\bibnamefont
  {Verma}}, \bibinfo {author} {\bibfnamefont {A.~A.}\ \bibnamefont {Franco}},\
  and\ \bibinfo {author} {\bibfnamefont {P.~P.}\ \bibnamefont {Mukherjee}},\
  }\href {https://doi.org/10.1149/2.0961915jes} {\bibfield  {journal} {\bibinfo
   {journal} {Journal of The Electrochemical Society}\ }\textbf {\bibinfo
  {volume} {166}},\ \bibinfo {pages} {A3852} (\bibinfo {year}
  {2019})}\BibitemShut {NoStop}%
\bibitem [{\citenamefont {Chandrasekaran}\ and\ \citenamefont
  {Fuller}(2011)}]{Chandrasekaran2011}%
  \BibitemOpen
  \bibfield  {author} {\bibinfo {author} {\bibfnamefont {R.}~\bibnamefont
  {Chandrasekaran}}\ and\ \bibinfo {author} {\bibfnamefont {T.~F.}\
  \bibnamefont {Fuller}},\ }\href {https://doi.org/10.1149/1.3589301}
  {\bibfield  {journal} {\bibinfo  {journal} {Journal of The Electrochemical
  Society}\ }\textbf {\bibinfo {volume} {158}},\ \bibinfo {pages} {A859}
  (\bibinfo {year} {2011})}\BibitemShut {NoStop}%
\bibitem [{\citenamefont {Sturm}\ \emph {et~al.}(2019)\citenamefont {Sturm},
  \citenamefont {Rheinfeld}, \citenamefont {Zilberman}, \citenamefont
  {Spingler}, \citenamefont {Kosch}, \citenamefont {Frie},\ and\ \citenamefont
  {Jossen}}]{Sturm2019}%
  \BibitemOpen
  \bibfield  {author} {\bibinfo {author} {\bibfnamefont {J.}~\bibnamefont
  {Sturm}}, \bibinfo {author} {\bibfnamefont {A.}~\bibnamefont {Rheinfeld}},
  \bibinfo {author} {\bibfnamefont {I.}~\bibnamefont {Zilberman}}, \bibinfo
  {author} {\bibfnamefont {F.}~\bibnamefont {Spingler}}, \bibinfo {author}
  {\bibfnamefont {S.}~\bibnamefont {Kosch}}, \bibinfo {author} {\bibfnamefont
  {F.}~\bibnamefont {Frie}},\ and\ \bibinfo {author} {\bibfnamefont
  {A.}~\bibnamefont {Jossen}},\ }\href
  {https://doi.org/10.1016/j.jpowsour.2018.11.043} {\bibfield  {journal}
  {\bibinfo  {journal} {Journal of Power Sources}\ }\textbf {\bibinfo {volume}
  {412}},\ \bibinfo {pages} {204} (\bibinfo {year} {2019})}\BibitemShut
  {NoStop}%
\bibitem [{\citenamefont {Lory}\ \emph {et~al.}(2020)\citenamefont {Lory},
  \citenamefont {Mathieu}, \citenamefont {Genies}, \citenamefont {Reynier},
  \citenamefont {Boulineau}, \citenamefont {Hong},\ and\ \citenamefont
  {Chandesris}}]{Lory2020}%
  \BibitemOpen
  \bibfield  {author} {\bibinfo {author} {\bibfnamefont {P.-F.}\ \bibnamefont
  {Lory}}, \bibinfo {author} {\bibfnamefont {B.}~\bibnamefont {Mathieu}},
  \bibinfo {author} {\bibfnamefont {S.}~\bibnamefont {Genies}}, \bibinfo
  {author} {\bibfnamefont {Y.}~\bibnamefont {Reynier}}, \bibinfo {author}
  {\bibfnamefont {A.}~\bibnamefont {Boulineau}}, \bibinfo {author}
  {\bibfnamefont {W.}~\bibnamefont {Hong}},\ and\ \bibinfo {author}
  {\bibfnamefont {M.}~\bibnamefont {Chandesris}},\ }\href
  {https://doi.org/10.1149/1945-7111/abaa69} {\bibfield  {journal} {\bibinfo
  {journal} {Journal of The Electrochemical Society}\ }\textbf {\bibinfo
  {volume} {167}},\ \bibinfo {pages} {120506} (\bibinfo {year}
  {2020})}\BibitemShut {NoStop}%
\bibitem [{\citenamefont {Wang}\ \emph {et~al.}(2017)\citenamefont {Wang},
  \citenamefont {Xiao},\ and\ \citenamefont {Huang}}]{Wang2017}%
  \BibitemOpen
  \bibfield  {author} {\bibinfo {author} {\bibfnamefont {M.}~\bibnamefont
  {Wang}}, \bibinfo {author} {\bibfnamefont {X.}~\bibnamefont {Xiao}},\ and\
  \bibinfo {author} {\bibfnamefont {X.}~\bibnamefont {Huang}},\ }\href
  {https://doi.org/10.1016/j.jpowsour.2017.02.037} {\bibfield  {journal}
  {\bibinfo  {journal} {Journal of Power Sources}\ }\textbf {\bibinfo {volume}
  {348}},\ \bibinfo {pages} {66} (\bibinfo {year} {2017})}\BibitemShut
  {NoStop}%
\bibitem [{\citenamefont {Appiah}\ \emph {et~al.}(2019)\citenamefont {Appiah},
  \citenamefont {Kim}, \citenamefont {Song}, \citenamefont {Ryou},\ and\
  \citenamefont {Lee}}]{Appiah2019}%
  \BibitemOpen
  \bibfield  {author} {\bibinfo {author} {\bibfnamefont {W.~A.}\ \bibnamefont
  {Appiah}}, \bibinfo {author} {\bibfnamefont {D.}~\bibnamefont {Kim}},
  \bibinfo {author} {\bibfnamefont {J.}~\bibnamefont {Song}}, \bibinfo {author}
  {\bibfnamefont {M.}~\bibnamefont {Ryou}},\ and\ \bibinfo {author}
  {\bibfnamefont {Y.~M.}\ \bibnamefont {Lee}},\ }\href
  {https://doi.org/10.1002/batt.201900019} {\bibfield  {journal} {\bibinfo
  {journal} {Batteries \& Supercaps}\ }\textbf {\bibinfo {volume} {2}},\
  \bibinfo {pages} {541} (\bibinfo {year} {2019})}\BibitemShut {NoStop}%
\bibitem [{\citenamefont {MacFarlane}\ \emph {et~al.}(2014)\citenamefont
  {MacFarlane}, \citenamefont {Tachikawa}, \citenamefont {Forsyth},
  \citenamefont {Pringle}, \citenamefont {Howlett}, \citenamefont {Elliott},
  \citenamefont {Davis}, \citenamefont {Watanabe}, \citenamefont {Simon},\ and\
  \citenamefont {Angell}}]{MacFarlane2014}%
  \BibitemOpen
  \bibfield  {author} {\bibinfo {author} {\bibfnamefont {D.~R.}\ \bibnamefont
  {MacFarlane}}, \bibinfo {author} {\bibfnamefont {N.}~\bibnamefont
  {Tachikawa}}, \bibinfo {author} {\bibfnamefont {M.}~\bibnamefont {Forsyth}},
  \bibinfo {author} {\bibfnamefont {J.~M.}\ \bibnamefont {Pringle}}, \bibinfo
  {author} {\bibfnamefont {P.~C.}\ \bibnamefont {Howlett}}, \bibinfo {author}
  {\bibfnamefont {G.~D.}\ \bibnamefont {Elliott}}, \bibinfo {author}
  {\bibfnamefont {J.~H.}\ \bibnamefont {Davis}}, \bibinfo {author}
  {\bibfnamefont {M.}~\bibnamefont {Watanabe}}, \bibinfo {author}
  {\bibfnamefont {P.}~\bibnamefont {Simon}},\ and\ \bibinfo {author}
  {\bibfnamefont {C.~A.}\ \bibnamefont {Angell}},\ }\href
  {https://doi.org/10.1039/C3EE42099J} {\bibfield  {journal} {\bibinfo
  {journal} {Energy \& Environmental Science}\ }\textbf {\bibinfo {volume}
  {7}},\ \bibinfo {pages} {232} (\bibinfo {year} {2014})}\BibitemShut {NoStop}%
\bibitem [{\citenamefont {Baranchugov}\ \emph {et~al.}(2007)\citenamefont
  {Baranchugov}, \citenamefont {Markevich}, \citenamefont {Pollak},
  \citenamefont {Salitra},\ and\ \citenamefont {Aurbach}}]{Baranchugov2007}%
  \BibitemOpen
  \bibfield  {author} {\bibinfo {author} {\bibfnamefont {V.}~\bibnamefont
  {Baranchugov}}, \bibinfo {author} {\bibfnamefont {E.}~\bibnamefont
  {Markevich}}, \bibinfo {author} {\bibfnamefont {E.}~\bibnamefont {Pollak}},
  \bibinfo {author} {\bibfnamefont {G.}~\bibnamefont {Salitra}},\ and\ \bibinfo
  {author} {\bibfnamefont {D.}~\bibnamefont {Aurbach}},\ }\href
  {https://doi.org/10.1016/j.elecom.2006.11.014} {\bibfield  {journal}
  {\bibinfo  {journal} {Electrochemistry Communications}\ }\textbf {\bibinfo
  {volume} {9}},\ \bibinfo {pages} {796} (\bibinfo {year} {2007})}\BibitemShut
  {NoStop}%
\bibitem [{\citenamefont {Tang}\ \emph {et~al.}(2022)\citenamefont {Tang},
  \citenamefont {Lv}, \citenamefont {Jiang}, \citenamefont {Zhou},\ and\
  \citenamefont {Liu}}]{Tang2022}%
  \BibitemOpen
  \bibfield  {author} {\bibinfo {author} {\bibfnamefont {X.}~\bibnamefont
  {Tang}}, \bibinfo {author} {\bibfnamefont {S.}~\bibnamefont {Lv}}, \bibinfo
  {author} {\bibfnamefont {K.}~\bibnamefont {Jiang}}, \bibinfo {author}
  {\bibfnamefont {G.}~\bibnamefont {Zhou}},\ and\ \bibinfo {author}
  {\bibfnamefont {X.}~\bibnamefont {Liu}},\ }\href
  {https://doi.org/10.1016/j.jpowsour.2022.231792} {\bibfield  {journal}
  {\bibinfo  {journal} {Journal of Power Sources}\ }\textbf {\bibinfo {volume}
  {542}},\ \bibinfo {pages} {231792} (\bibinfo {year} {2022})}\BibitemShut
  {NoStop}%
\bibitem [{\citenamefont {Chakrapani}\ \emph {et~al.}(2011)\citenamefont
  {Chakrapani}, \citenamefont {Rusli}, \citenamefont {Filler},\ and\
  \citenamefont {Kohl}}]{Chakrapani2011}%
  \BibitemOpen
  \bibfield  {author} {\bibinfo {author} {\bibfnamefont {V.}~\bibnamefont
  {Chakrapani}}, \bibinfo {author} {\bibfnamefont {F.}~\bibnamefont {Rusli}},
  \bibinfo {author} {\bibfnamefont {M.~A.}\ \bibnamefont {Filler}},\ and\
  \bibinfo {author} {\bibfnamefont {P.~A.}\ \bibnamefont {Kohl}},\ }\href
  {https://doi.org/10.1021/jp207605w} {\bibfield  {journal} {\bibinfo
  {journal} {The Journal of Physical Chemistry C}\ }\textbf {\bibinfo {volume}
  {115}},\ \bibinfo {pages} {22048} (\bibinfo {year} {2011})}\BibitemShut
  {NoStop}%
\bibitem [{\citenamefont {Kim}\ \emph {et~al.}(2017)\citenamefont {Kim},
  \citenamefont {Kennedy}, \citenamefont {Brandon}, \citenamefont {Geaney},
  \citenamefont {Ryan}, \citenamefont {Passerini},\ and\ \citenamefont
  {Appetecchi}}]{Kim2017}%
  \BibitemOpen
  \bibfield  {author} {\bibinfo {author} {\bibfnamefont {G.~T.}\ \bibnamefont
  {Kim}}, \bibinfo {author} {\bibfnamefont {T.}~\bibnamefont {Kennedy}},
  \bibinfo {author} {\bibfnamefont {M.}~\bibnamefont {Brandon}}, \bibinfo
  {author} {\bibfnamefont {H.}~\bibnamefont {Geaney}}, \bibinfo {author}
  {\bibfnamefont {K.~M.}\ \bibnamefont {Ryan}}, \bibinfo {author}
  {\bibfnamefont {S.}~\bibnamefont {Passerini}},\ and\ \bibinfo {author}
  {\bibfnamefont {G.~B.}\ \bibnamefont {Appetecchi}},\ }\href
  {https://doi.org/10.1021/acsnano.7b01705} {\bibfield  {journal} {\bibinfo
  {journal} {ACS Nano}\ }\textbf {\bibinfo {volume} {11}},\ \bibinfo {pages}
  {5933} (\bibinfo {year} {2017})}\BibitemShut {NoStop}%
\bibitem [{\citenamefont {Kerr}\ \emph {et~al.}(2017)\citenamefont {Kerr},
  \citenamefont {Mazouzi}, \citenamefont {Eftekharnia}, \citenamefont
  {Lestriez}, \citenamefont {Dupré}, \citenamefont {Forsyth}, \citenamefont
  {Guyomard},\ and\ \citenamefont {Howlett}}]{Kerr2017}%
  \BibitemOpen
  \bibfield  {author} {\bibinfo {author} {\bibfnamefont {R.}~\bibnamefont
  {Kerr}}, \bibinfo {author} {\bibfnamefont {D.}~\bibnamefont {Mazouzi}},
  \bibinfo {author} {\bibfnamefont {M.}~\bibnamefont {Eftekharnia}}, \bibinfo
  {author} {\bibfnamefont {B.}~\bibnamefont {Lestriez}}, \bibinfo {author}
  {\bibfnamefont {N.}~\bibnamefont {Dupré}}, \bibinfo {author} {\bibfnamefont
  {M.}~\bibnamefont {Forsyth}}, \bibinfo {author} {\bibfnamefont
  {D.}~\bibnamefont {Guyomard}},\ and\ \bibinfo {author} {\bibfnamefont
  {P.~C.}\ \bibnamefont {Howlett}},\ }\href
  {https://doi.org/10.1021/acsenergylett.7b00403} {\bibfield  {journal}
  {\bibinfo  {journal} {ACS Energy Letters}\ }\textbf {\bibinfo {volume} {2}},\
  \bibinfo {pages} {1804} (\bibinfo {year} {2017})}\BibitemShut {NoStop}%
\bibitem [{\citenamefont {Stokes}\ \emph {et~al.}(2020)\citenamefont {Stokes},
  \citenamefont {Kennedy}, \citenamefont {Kim}, \citenamefont {Geaney},
  \citenamefont {Storan}, \citenamefont {Laffir}, \citenamefont {Appetecchi},
  \citenamefont {Passerini},\ and\ \citenamefont {Ryan}}]{Stokes2020}%
  \BibitemOpen
  \bibfield  {author} {\bibinfo {author} {\bibfnamefont {K.}~\bibnamefont
  {Stokes}}, \bibinfo {author} {\bibfnamefont {T.}~\bibnamefont {Kennedy}},
  \bibinfo {author} {\bibfnamefont {G.-T.}\ \bibnamefont {Kim}}, \bibinfo
  {author} {\bibfnamefont {H.}~\bibnamefont {Geaney}}, \bibinfo {author}
  {\bibfnamefont {D.}~\bibnamefont {Storan}}, \bibinfo {author} {\bibfnamefont
  {F.}~\bibnamefont {Laffir}}, \bibinfo {author} {\bibfnamefont {G.~B.}\
  \bibnamefont {Appetecchi}}, \bibinfo {author} {\bibfnamefont
  {S.}~\bibnamefont {Passerini}},\ and\ \bibinfo {author} {\bibfnamefont
  {K.~M.}\ \bibnamefont {Ryan}},\ }\href
  {https://doi.org/10.1021/acs.nanolett.0c01774} {\bibfield  {journal}
  {\bibinfo  {journal} {Nano Letters}\ }\textbf {\bibinfo {volume} {20}},\
  \bibinfo {pages} {7011} (\bibinfo {year} {2020})}\BibitemShut {NoStop}%
\bibitem [{\citenamefont {Domi}\ \emph {et~al.}(2022)\citenamefont {Domi},
  \citenamefont {Usui}, \citenamefont {Hirosawa}, \citenamefont {Sugimoto},
  \citenamefont {Nakano}, \citenamefont {Ando},\ and\ \citenamefont
  {Sakaguchi}}]{Domi2022}%
  \BibitemOpen
  \bibfield  {author} {\bibinfo {author} {\bibfnamefont {Y.}~\bibnamefont
  {Domi}}, \bibinfo {author} {\bibfnamefont {H.}~\bibnamefont {Usui}}, \bibinfo
  {author} {\bibfnamefont {T.}~\bibnamefont {Hirosawa}}, \bibinfo {author}
  {\bibfnamefont {K.}~\bibnamefont {Sugimoto}}, \bibinfo {author}
  {\bibfnamefont {T.}~\bibnamefont {Nakano}}, \bibinfo {author} {\bibfnamefont
  {A.}~\bibnamefont {Ando}},\ and\ \bibinfo {author} {\bibfnamefont
  {H.}~\bibnamefont {Sakaguchi}},\ }\href
  {https://doi.org/10.1021/acsomega.2c00947} {\bibfield  {journal} {\bibinfo
  {journal} {ACS Omega}\ }\textbf {\bibinfo {volume} {7}},\ \bibinfo {pages}
  {15846} (\bibinfo {year} {2022})}\BibitemShut {NoStop}%
\bibitem [{\citenamefont {Karimi}\ \emph {et~al.}(2023)\citenamefont {Karimi},
  \citenamefont {Zarrabeitia}, \citenamefont {Geaney}, \citenamefont {Ryan},
  \citenamefont {Iliev}, \citenamefont {Schubert}, \citenamefont {Varzi},\ and\
  \citenamefont {Passerini}}]{Karimi2023}%
  \BibitemOpen
  \bibfield  {author} {\bibinfo {author} {\bibfnamefont {N.}~\bibnamefont
  {Karimi}}, \bibinfo {author} {\bibfnamefont {M.}~\bibnamefont {Zarrabeitia}},
  \bibinfo {author} {\bibfnamefont {H.}~\bibnamefont {Geaney}}, \bibinfo
  {author} {\bibfnamefont {K.~M.}\ \bibnamefont {Ryan}}, \bibinfo {author}
  {\bibfnamefont {B.}~\bibnamefont {Iliev}}, \bibinfo {author} {\bibfnamefont
  {T.~J.}\ \bibnamefont {Schubert}}, \bibinfo {author} {\bibfnamefont
  {A.}~\bibnamefont {Varzi}},\ and\ \bibinfo {author} {\bibfnamefont
  {S.}~\bibnamefont {Passerini}},\ }\href
  {https://doi.org/10.1016/j.jpowsour.2022.232621} {\bibfield  {journal}
  {\bibinfo  {journal} {Journal of Power Sources}\ }\textbf {\bibinfo {volume}
  {558}},\ \bibinfo {pages} {232621} (\bibinfo {year} {2023})}\BibitemShut
  {NoStop}%
\bibitem [{\citenamefont {Bazant}\ \emph {et~al.}(2011)\citenamefont {Bazant},
  \citenamefont {Storey},\ and\ \citenamefont {Kornyshev}}]{Bazant2011}%
  \BibitemOpen
  \bibfield  {author} {\bibinfo {author} {\bibfnamefont {M.~Z.}\ \bibnamefont
  {Bazant}}, \bibinfo {author} {\bibfnamefont {B.~D.}\ \bibnamefont {Storey}},\
  and\ \bibinfo {author} {\bibfnamefont {A.~A.}\ \bibnamefont {Kornyshev}},\
  }\href {https://doi.org/10.1103/PhysRevLett.106.046102} {\bibfield  {journal}
  {\bibinfo  {journal} {Physical Review Letters}\ }\textbf {\bibinfo {volume}
  {106}},\ \bibinfo {pages} {046102} (\bibinfo {year} {2011})}\BibitemShut
  {NoStop}%
\bibitem [{\citenamefont {Yochelis}\ \emph {et~al.}(2015)\citenamefont
  {Yochelis}, \citenamefont {Singh},\ and\ \citenamefont
  {Visoly-Fisher}}]{Yochelis2015}%
  \BibitemOpen
  \bibfield  {author} {\bibinfo {author} {\bibfnamefont {A.}~\bibnamefont
  {Yochelis}}, \bibinfo {author} {\bibfnamefont {M.~B.}\ \bibnamefont
  {Singh}},\ and\ \bibinfo {author} {\bibfnamefont {I.}~\bibnamefont
  {Visoly-Fisher}},\ }\href {https://doi.org/10.1021/acs.chemmater.5b00780}
  {\bibfield  {journal} {\bibinfo  {journal} {Chemistry of Materials}\ }\textbf
  {\bibinfo {volume} {27}},\ \bibinfo {pages} {4169} (\bibinfo {year}
  {2015})}\BibitemShut {NoStop}%
\bibitem [{\citenamefont {Gavish}\ and\ \citenamefont
  {Yochelis}(2016)}]{Gavish2016}%
  \BibitemOpen
  \bibfield  {author} {\bibinfo {author} {\bibfnamefont {N.}~\bibnamefont
  {Gavish}}\ and\ \bibinfo {author} {\bibfnamefont {A.}~\bibnamefont
  {Yochelis}},\ }\href {https://doi.org/10.1021/acs.jpclett.6b00370} {\bibfield
   {journal} {\bibinfo  {journal} {The Journal of Physical Chemistry Letters}\
  }\textbf {\bibinfo {volume} {7}},\ \bibinfo {pages} {1121} (\bibinfo {year}
  {2016})}\BibitemShut {NoStop}%
\bibitem [{\citenamefont {Gavish}\ \emph {et~al.}(2017)\citenamefont {Gavish},
  \citenamefont {Elad},\ and\ \citenamefont {Yochelis}}]{Gavish2017}%
  \BibitemOpen
  \bibfield  {author} {\bibinfo {author} {\bibfnamefont {N.}~\bibnamefont
  {Gavish}}, \bibinfo {author} {\bibfnamefont {D.}~\bibnamefont {Elad}},\ and\
  \bibinfo {author} {\bibfnamefont {A.}~\bibnamefont {Yochelis}},\ }\href
  {https://doi.org/10.1021/acs.jpclett.7b03048} {\bibfield  {journal} {\bibinfo
   {journal} {Journal of Physical Chemistry Letters}\ }\textbf {\bibinfo
  {volume} {9}},\ \bibinfo {pages} {36} (\bibinfo {year} {2017})}\BibitemShut
  {NoStop}%
\bibitem [{\citenamefont {Bier}\ \emph {et~al.}(2017)\citenamefont {Bier},
  \citenamefont {Gavish}, \citenamefont {Uecker},\ and\ \citenamefont
  {Yochelis}}]{Bier2017}%
  \BibitemOpen
  \bibfield  {author} {\bibinfo {author} {\bibfnamefont {S.}~\bibnamefont
  {Bier}}, \bibinfo {author} {\bibfnamefont {N.}~\bibnamefont {Gavish}},
  \bibinfo {author} {\bibfnamefont {H.}~\bibnamefont {Uecker}},\ and\ \bibinfo
  {author} {\bibfnamefont {A.}~\bibnamefont {Yochelis}},\ }\href
  {https://doi.org/10.1103/PhysRevE.95.060201} {\bibfield  {journal} {\bibinfo
  {journal} {Physical Review E}\ }\textbf {\bibinfo {volume} {95}},\ \bibinfo
  {pages} {060201} (\bibinfo {year} {2017})}\BibitemShut {NoStop}%
\bibitem [{\citenamefont {Monroe}\ and\ \citenamefont
  {Delacourt}(2013)}]{Monroe2013}%
  \BibitemOpen
  \bibfield  {author} {\bibinfo {author} {\bibfnamefont {C.~W.}\ \bibnamefont
  {Monroe}}\ and\ \bibinfo {author} {\bibfnamefont {C.}~\bibnamefont
  {Delacourt}},\ }\href {https://doi.org/10.1016/j.electacta.2013.10.006}
  {\bibfield  {journal} {\bibinfo  {journal} {Electrochimica Acta}\ }\textbf
  {\bibinfo {volume} {114}},\ \bibinfo {pages} {649} (\bibinfo {year}
  {2013})}\BibitemShut {NoStop}%
\bibitem [{\citenamefont {Liu}\ and\ \citenamefont {Monroe}(2014)}]{LiuJ2014}%
  \BibitemOpen
  \bibfield  {author} {\bibinfo {author} {\bibfnamefont {J.}~\bibnamefont
  {Liu}}\ and\ \bibinfo {author} {\bibfnamefont {C.~W.}\ \bibnamefont
  {Monroe}},\ }\href {https://doi.org/10.1016/j.electacta.2014.05.009}
  {\bibfield  {journal} {\bibinfo  {journal} {Electrochimica Acta}\ }\textbf
  {\bibinfo {volume} {135}},\ \bibinfo {pages} {447} (\bibinfo {year}
  {2014})}\BibitemShut {NoStop}%
\bibitem [{\citenamefont {Goyal}\ and\ \citenamefont
  {Monroe}(2017)}]{Goyal2017}%
  \BibitemOpen
  \bibfield  {author} {\bibinfo {author} {\bibfnamefont {P.}~\bibnamefont
  {Goyal}}\ and\ \bibinfo {author} {\bibfnamefont {C.~W.}\ \bibnamefont
  {Monroe}},\ }\href {https://doi.org/10.1149/2.0611711jes} {\bibfield
  {journal} {\bibinfo  {journal} {Journal of The Electrochemical Society}\
  }\textbf {\bibinfo {volume} {164}},\ \bibinfo {pages} {E3647} (\bibinfo
  {year} {2017})}\BibitemShut {NoStop}%
\bibitem [{\citenamefont {Schammer}\ \emph {et~al.}(2021)\citenamefont
  {Schammer}, \citenamefont {Horstmann},\ and\ \citenamefont
  {Latz}}]{Schammer2020theory}%
  \BibitemOpen
  \bibfield  {author} {\bibinfo {author} {\bibfnamefont {M.}~\bibnamefont
  {Schammer}}, \bibinfo {author} {\bibfnamefont {B.}~\bibnamefont
  {Horstmann}},\ and\ \bibinfo {author} {\bibfnamefont {A.}~\bibnamefont
  {Latz}},\ }\href {https://doi.org/10.1149/1945-7111/abdddf} {\bibfield
  {journal} {\bibinfo  {journal} {Journal of the Electrochemical Society}\
  }\textbf {\bibinfo {volume} {168}},\ \bibinfo {pages} {026511} (\bibinfo
  {year} {2021})}\BibitemShut {NoStop}%
\bibitem [{\citenamefont {Hoffmann}\ \emph {et~al.}(2018)\citenamefont
  {Hoffmann}, \citenamefont {Pulletikurthi}, \citenamefont {Carstens},
  \citenamefont {Lahiri}, \citenamefont {Borodin}, \citenamefont {Schammer},
  \citenamefont {Horstmann}, \citenamefont {Latz},\ and\ \citenamefont
  {Endres}}]{Hoffmann2018}%
  \BibitemOpen
  \bibfield  {author} {\bibinfo {author} {\bibfnamefont {V.}~\bibnamefont
  {Hoffmann}}, \bibinfo {author} {\bibfnamefont {G.}~\bibnamefont
  {Pulletikurthi}}, \bibinfo {author} {\bibfnamefont {T.}~\bibnamefont
  {Carstens}}, \bibinfo {author} {\bibfnamefont {A.}~\bibnamefont {Lahiri}},
  \bibinfo {author} {\bibfnamefont {A.}~\bibnamefont {Borodin}}, \bibinfo
  {author} {\bibfnamefont {M.}~\bibnamefont {Schammer}}, \bibinfo {author}
  {\bibfnamefont {B.}~\bibnamefont {Horstmann}}, \bibinfo {author}
  {\bibfnamefont {A.}~\bibnamefont {Latz}},\ and\ \bibinfo {author}
  {\bibfnamefont {F.}~\bibnamefont {Endres}},\ }\href
  {https://doi.org/10.1039/C7CP08243F} {\bibfield  {journal} {\bibinfo
  {journal} {Physical Chemistry Chemical Physics}\ }\textbf {\bibinfo {volume}
  {20}},\ \bibinfo {pages} {4760} (\bibinfo {year} {2018})}\BibitemShut
  {NoStop}%
\bibitem [{\citenamefont {Schammer}\ \emph {et~al.}(2022)\citenamefont
  {Schammer}, \citenamefont {Latz},\ and\ \citenamefont
  {Horstmann}}]{Schammer2021role}%
  \BibitemOpen
  \bibfield  {author} {\bibinfo {author} {\bibfnamefont {M.}~\bibnamefont
  {Schammer}}, \bibinfo {author} {\bibfnamefont {A.}~\bibnamefont {Latz}},\
  and\ \bibinfo {author} {\bibfnamefont {B.}~\bibnamefont {Horstmann}},\ }\href
  {https://doi.org/10.1021/acs.jpcb.2c00215} {\bibfield  {journal} {\bibinfo
  {journal} {The Journal of Physical Chemistry B}\ }\textbf {\bibinfo {volume}
  {126}},\ \bibinfo {pages} {2761} (\bibinfo {year} {2022})}\BibitemShut
  {NoStop}%
\bibitem [{\citenamefont {Latz}\ and\ \citenamefont {Zausch}(2015)}]{Latz2015}%
  \BibitemOpen
  \bibfield  {author} {\bibinfo {author} {\bibfnamefont {A.}~\bibnamefont
  {Latz}}\ and\ \bibinfo {author} {\bibfnamefont {J.}~\bibnamefont {Zausch}},\
  }\href {https://doi.org/10.3762/bjnano.6.102} {\bibfield  {journal} {\bibinfo
   {journal} {Beilstein Journal of Nanotechnology}\ }\textbf {\bibinfo {volume}
  {6}},\ \bibinfo {pages} {987} (\bibinfo {year} {2015})}\BibitemShut {NoStop}%
\bibitem [{\citenamefont {Newman}\ and\ \citenamefont
  {Tiedemann}(1975)}]{Newman1975}%
  \BibitemOpen
  \bibfield  {author} {\bibinfo {author} {\bibfnamefont {J.}~\bibnamefont
  {Newman}}\ and\ \bibinfo {author} {\bibfnamefont {W.}~\bibnamefont
  {Tiedemann}},\ }\href {https://doi.org/10.1002/aic.690210103} {\bibfield
  {journal} {\bibinfo  {journal} {AIChE Journal}\ }\textbf {\bibinfo {volume}
  {21}},\ \bibinfo {pages} {25} (\bibinfo {year} {1975})}\BibitemShut {NoStop}%
\bibitem [{\citenamefont {Newman}\ and\ \citenamefont
  {Thomas-Alyea}(2012)}]{Newman2012}%
  \BibitemOpen
  \bibfield  {author} {\bibinfo {author} {\bibfnamefont {J.}~\bibnamefont
  {Newman}}\ and\ \bibinfo {author} {\bibfnamefont {K.~E.}\ \bibnamefont
  {Thomas-Alyea}},\ }\href@noop {} {\emph {\bibinfo {title} {Electrochemical
  systems}}}\ (\bibinfo  {publisher} {John Wiley \& Sons},\ \bibinfo {year}
  {2012})\BibitemShut {NoStop}%
\bibitem [{\citenamefont {Danner}\ \emph {et~al.}(2016)\citenamefont {Danner},
  \citenamefont {Singh}, \citenamefont {Hein}, \citenamefont {Kaiser},
  \citenamefont {Hahn},\ and\ \citenamefont {Latz}}]{Danner2016}%
  \BibitemOpen
  \bibfield  {author} {\bibinfo {author} {\bibfnamefont {T.}~\bibnamefont
  {Danner}}, \bibinfo {author} {\bibfnamefont {M.}~\bibnamefont {Singh}},
  \bibinfo {author} {\bibfnamefont {S.}~\bibnamefont {Hein}}, \bibinfo {author}
  {\bibfnamefont {J.}~\bibnamefont {Kaiser}}, \bibinfo {author} {\bibfnamefont
  {H.}~\bibnamefont {Hahn}},\ and\ \bibinfo {author} {\bibfnamefont
  {A.}~\bibnamefont {Latz}},\ }\href
  {https://doi.org/10.1016/j.jpowsour.2016.09.143} {\bibfield  {journal}
  {\bibinfo  {journal} {Journal of Power Sources}\ }\textbf {\bibinfo {volume}
  {334}},\ \bibinfo {pages} {191} (\bibinfo {year} {2016})}\BibitemShut
  {NoStop}%
\bibitem [{\citenamefont {Lorenz}\ and\ \citenamefont
  {Schönhoff}(2023)}]{Lorenz2023}%
  \BibitemOpen
  \bibfield  {author} {\bibinfo {author} {\bibfnamefont {M.}~\bibnamefont
  {Lorenz}}\ and\ \bibinfo {author} {\bibfnamefont {M.}~\bibnamefont
  {Schönhoff}},\ }\href@noop {} {\bibfield  {journal} {\bibinfo  {journal}
  {The Journal of Physical Chemistry B}\ } (\bibinfo {year} {2023})},\ \bibinfo
  {note} {[submitted]}\BibitemShut {NoStop}%
\bibitem [{\citenamefont {Lorenz}\ \emph {et~al.}(2022)\citenamefont {Lorenz},
  \citenamefont {Kilchert}, \citenamefont {Nürnberg}, \citenamefont
  {Schammer}, \citenamefont {Latz}, \citenamefont {Horstmann},\ and\
  \citenamefont {Schönhoff}}]{Lorenz2022}%
  \BibitemOpen
  \bibfield  {author} {\bibinfo {author} {\bibfnamefont {M.}~\bibnamefont
  {Lorenz}}, \bibinfo {author} {\bibfnamefont {F.}~\bibnamefont {Kilchert}},
  \bibinfo {author} {\bibfnamefont {P.}~\bibnamefont {Nürnberg}}, \bibinfo
  {author} {\bibfnamefont {M.}~\bibnamefont {Schammer}}, \bibinfo {author}
  {\bibfnamefont {A.}~\bibnamefont {Latz}}, \bibinfo {author} {\bibfnamefont
  {B.}~\bibnamefont {Horstmann}},\ and\ \bibinfo {author} {\bibfnamefont
  {M.}~\bibnamefont {Schönhoff}},\ }\href
  {https://doi.org/10.1021/acs.jpclett.2c02398} {\bibfield  {journal} {\bibinfo
   {journal} {The Journal of Physical Chemistry Letters}\ }\textbf {\bibinfo
  {volume} {13}},\ \bibinfo {pages} {8761} (\bibinfo {year}
  {2022})}\BibitemShut {NoStop}%
\bibitem [{\citenamefont {Latz}\ and\ \citenamefont {Zausch}(2013)}]{Latz2013}%
  \BibitemOpen
  \bibfield  {author} {\bibinfo {author} {\bibfnamefont {A.}~\bibnamefont
  {Latz}}\ and\ \bibinfo {author} {\bibfnamefont {J.}~\bibnamefont {Zausch}},\
  }\href {https://doi.org/10.1016/j.electacta.2013.06.043} {\bibfield
  {journal} {\bibinfo  {journal} {Electrochimica Acta}\ }\textbf {\bibinfo
  {volume} {110}},\ \bibinfo {pages} {358} (\bibinfo {year}
  {2013})}\BibitemShut {NoStop}%
\bibitem [{\citenamefont {Stokes}\ \emph {et~al.}(2019)\citenamefont {Stokes},
  \citenamefont {Geaney}, \citenamefont {Sheehan}, \citenamefont {Borsa},\ and\
  \citenamefont {Ryan}}]{Stokes2019_2}%
  \BibitemOpen
  \bibfield  {author} {\bibinfo {author} {\bibfnamefont {K.}~\bibnamefont
  {Stokes}}, \bibinfo {author} {\bibfnamefont {H.}~\bibnamefont {Geaney}},
  \bibinfo {author} {\bibfnamefont {M.}~\bibnamefont {Sheehan}}, \bibinfo
  {author} {\bibfnamefont {D.}~\bibnamefont {Borsa}},\ and\ \bibinfo {author}
  {\bibfnamefont {K.~M.}\ \bibnamefont {Ryan}},\ }\href
  {https://doi.org/10.1021/acs.nanolett.9b03664} {\bibfield  {journal}
  {\bibinfo  {journal} {Nano Letters}\ }\textbf {\bibinfo {volume} {19}},\
  \bibinfo {pages} {8829} (\bibinfo {year} {2019})}\BibitemShut {NoStop}%
\bibitem [{\citenamefont {Kilchert}\ \emph {et~al.}(2023)\citenamefont
  {Kilchert}, \citenamefont {Lorenz}, \citenamefont {Schammer}, \citenamefont
  {Nürnberg}, \citenamefont {Schönhoff}, \citenamefont {Latz},\ and\
  \citenamefont {Horstmann}}]{Kilchert2023}%
  \BibitemOpen
  \bibfield  {author} {\bibinfo {author} {\bibfnamefont {F.}~\bibnamefont
  {Kilchert}}, \bibinfo {author} {\bibfnamefont {M.}~\bibnamefont {Lorenz}},
  \bibinfo {author} {\bibfnamefont {M.}~\bibnamefont {Schammer}}, \bibinfo
  {author} {\bibfnamefont {P.}~\bibnamefont {Nürnberg}}, \bibinfo {author}
  {\bibfnamefont {M.}~\bibnamefont {Schönhoff}}, \bibinfo {author}
  {\bibfnamefont {A.}~\bibnamefont {Latz}},\ and\ \bibinfo {author}
  {\bibfnamefont {B.}~\bibnamefont {Horstmann}},\ }\href
  {https://doi.org/10.1039/D2CP04423D} {\bibfield  {journal} {\bibinfo
  {journal} {Physical Chemistry Chemical Physics}\ }\textbf {\bibinfo {volume}
  {25}},\ \bibinfo {pages} {25965} (\bibinfo {year} {2023})}\BibitemShut
  {NoStop}%
\bibitem [{\citenamefont {Pan}\ \emph {et~al.}(2019)\citenamefont {Pan},
  \citenamefont {Zou}, \citenamefont {Canova}, \citenamefont {Zhu},\ and\
  \citenamefont {Kim}}]{Pan2019}%
  \BibitemOpen
  \bibfield  {author} {\bibinfo {author} {\bibfnamefont {K.}~\bibnamefont
  {Pan}}, \bibinfo {author} {\bibfnamefont {F.}~\bibnamefont {Zou}}, \bibinfo
  {author} {\bibfnamefont {M.}~\bibnamefont {Canova}}, \bibinfo {author}
  {\bibfnamefont {Y.}~\bibnamefont {Zhu}},\ and\ \bibinfo {author}
  {\bibfnamefont {J.-H.}\ \bibnamefont {Kim}},\ }\href
  {https://doi.org/10.1016/j.jpowsour.2018.12.010} {\bibfield  {journal}
  {\bibinfo  {journal} {Journal of Power Sources}\ }\textbf {\bibinfo {volume}
  {413}},\ \bibinfo {pages} {20} (\bibinfo {year} {2019})}\BibitemShut
  {NoStop}%
\bibitem [{\citenamefont {Waldmann}\ \emph {et~al.}(2018)\citenamefont
  {Waldmann}, \citenamefont {Hogg},\ and\ \citenamefont
  {Wohlfahrt-Mehrens}}]{Waldmann2018}%
  \BibitemOpen
  \bibfield  {author} {\bibinfo {author} {\bibfnamefont {T.}~\bibnamefont
  {Waldmann}}, \bibinfo {author} {\bibfnamefont {B.-I.}\ \bibnamefont {Hogg}},\
  and\ \bibinfo {author} {\bibfnamefont {M.}~\bibnamefont
  {Wohlfahrt-Mehrens}},\ }\href
  {https://doi.org/10.1016/j.jpowsour.2018.02.063} {\bibfield  {journal}
  {\bibinfo  {journal} {Journal of Power Sources}\ }\textbf {\bibinfo {volume}
  {384}},\ \bibinfo {pages} {107} (\bibinfo {year} {2018})}\BibitemShut
  {NoStop}%
\bibitem [{\citenamefont {Wang}\ \emph {et~al.}(2016)\citenamefont {Wang},
  \citenamefont {Xiao},\ and\ \citenamefont {Huang}}]{Wang2016}%
  \BibitemOpen
  \bibfield  {author} {\bibinfo {author} {\bibfnamefont {M.}~\bibnamefont
  {Wang}}, \bibinfo {author} {\bibfnamefont {X.}~\bibnamefont {Xiao}},\ and\
  \bibinfo {author} {\bibfnamefont {X.}~\bibnamefont {Huang}},\ }\href
  {https://doi.org/10.1016/j.jpowsour.2015.12.082} {\bibfield  {journal}
  {\bibinfo  {journal} {Journal of Power Sources}\ }\textbf {\bibinfo {volume}
  {307}},\ \bibinfo {pages} {77} (\bibinfo {year} {2016})}\BibitemShut
  {NoStop}%
\end{thebibliography}%

\end{document}